\documentclass[fleqn,usenatbib]{mnras}
\usepackage{newtxtext,newtxmath}

\usepackage[T1]{fontenc}
\usepackage{ae,aecompl}

\usepackage{graphicx}	% Including figure files
\usepackage{amsmath}	% Advanced maths commands
\usepackage{xspace}
\usepackage{amsfonts,textcomp}
\usepackage[usenames]{color}
\usepackage{times}
\usepackage{hyperref}
\usepackage[utf8]{inputenc}
\usepackage{natbib}
\newcommand{\be}{\begin{equation}}
\newcommand{\ee}{\end{equation}}

\newcommand{\msun}{\rm M_{\sun}}

\newcommand{\nh}{n_{\rm H}}

\definecolor{darkgreen}{rgb}{0.0,0.5,0.0}
\definecolor{darkred}{rgb}{0.5,0.0,0.0}
\definecolor{brown}{rgb}{0.65,.16,0.246}
\definecolor{grey}{rgb}{0.4,0.5,0.6}

\title[HDGAS-II; Neutral Atomic to Molecular Gas Phases]{Hydrodynamic simulations of the Disc of Gas Around Supermassive black holes (HDGAS) -II; The transition from neutral atomic to molecular gas phases}
\author[M. Raouf et al.]{
Mojtaba~Raouf,$^{1,2,7}$\thanks{E-mail: *mojtaba.raouf@gmail.com, raouf@strw.leidenuniv.nl} 
Serena Viti,$^{1,3,4}$
Reihaneh Karimi,$^{2}$
Alexander J. Richings,$^{5,6}$\\
$^{1}$ Leiden Observatory, Leiden University, P.O. Box 9513, 2300 RA Leiden, Netherlands\\
$^{2}$  School of Astronomy, Institute for Research in Fundamental Sciences (IPM), Tehran, 19395-5746, Iran \\
$^{3}$ Transdisciplinary Research Area (TRA) `Matter'/Argelander-Institut für Astronomie, University of Bonn, Germany\\
$^{4}$ Department of Physics and Astronomy, University College London, Gower Street, WC1E 6BT, London, UK\\
$^{5}$ E. A. Milne Centre for Astrophysics, Department of Physics and Mathematics, University of Hull, Cottingham Road, Hull, HU6 7RX, UK\\
$^{6}$ DAIM, University of Hull, Cottingham Road, Hull, HU6 7RX, UK\\
$^{7}$ LUNEX EuroMoonMars EuroSpaceHub, SBIC Space Business Innovation Centre, Kapteynstraat 1, 2201 BB Noordwijk,
The Netherlands
}
\begin{document}
\label{firstpage}
\pagerange{\pageref{firstpage}--\pageref{lastpage}}
\maketitle
%####################################################################
% Abstract of the paper
\begin{abstract}
We use HDGAS hydrodynamic simulations to study the impact of active galactic nucleus (AGN) feedback on the conversion of atomic-gas to molecular-gas within the circumnuclear-disc (CND) of a typical AGN-dominated galaxy. The comparison of CI, CII, and CO line intensities and their ratios in the HDGAS post-processing radiative-transfer analysis reveals the complex interplay between AGN-activity, cold molecular gas properties, and the physical processes governing the evolution of star-formation in galaxies.
Our results demonstrate that the CI/CO intensity ratio serves as a reliable indicator of the atomic-to-molecular gas transition. We present the probability distribution function (PDF) and abundance trends of various metal species related to molecular H$2$ gas, highlighting differences in clumpiness and intensity maps between AGN feedback and NoAGN models. The profile of the integrated intensity (moment-0) maps shows that the AGN-feedback model exhibits a lower CI/CO intensity ratio in the vicinity of the supermassive black hole (< 50 pc),  indicating a smaller atomic-gas abundance
and the presence of positive AGN-feedback. Our simulations have successfully predicted the presence of faint-CO emissions extending to larger radii from the galactic center. We also explore the relationships between CII/CO and CI/CII intensity ratios, as well as the ratios versus CO intensity, which provides insights into the "CO-dark" issues.
One notable feature in the later time-scale of the AGN model is the presence of a "CO-dark" region, where the intensity of CO emission ($\rm I_{CO}$) is depleted relative to the H$_2$ column density ($N_{\rm H_2}$) compared to the NoAGN model. 

\end{abstract}


\section{Introduction}
Molecular gas is both a product and a driver of star formation \citep[e.g.][]{Kennicutt2012}. It forms from atomic hydrogen in cold and shielded regions of the interstellar medium (ISM), where it can collapse and fragment to form stars. 
It impacts the star formation process by determining the physical and chemical conditions of the ISM, which can be examined through the observation of different atomic and molecular gas tracers. These tracers are employed to investigate the galactic disc and circumnuclear disc of galaxies \citep{Kamenetzky2011,Viti2014,Combes2014,Takano2014,Salvestrini2022}.

Atoms can serve as tracers of relatively dense molecular gas regions in different galaxies. Examples of such atoms include neutral atomic carbon (CI) and singly ionized carbon (CII) \citep{Papadopoulos2004, Bell2007,Glover2015,Bisbas2017}. 
The impact of cosmic rays (CRs) on the abundance and distribution of CO, CI, and \text{CII} within molecular clouds of star-forming galaxies has been the subject of recent investigations. \cite{Glover2015} explored the utility of CI emission in turbulent molecular clouds for investigating their structure and dynamics. Through numerical simulations of chemical, thermal, and dynamical evolution, the authors modeled CI emission under various cloud properties and observational conditions. The findings indicated widespread CI emission within the cloud due to turbulence-induced density variations, enabling radiation to penetrate deeply. The CI emission was demonstrated to accurately trace the column density of the cloud across a broad range of visual extinctions, surpassing the reliability of CO for low-extinction regions. The study further revealed that CI excitation temperatures generally remained lower than the kinetic temperatures, suggesting that the carbon atoms were not in thermal equilibrium with the gas. The authors discussed techniques for estimating CI excitation temperatures and atomic carbon column density based on CI line observations, while considering potential estimation errors. 
Other studies revealed that elevated CR ionization rates can lead to a substantial reduction in CO abundance, thereby complicating the tracing of $H_2$ mass and cloud structure. 
Additionally, CI and \text{CII} were found to become more prevalent and dominant at higher CR ionization rates \citep{Bisbas2017,Izumi2018}. 
\begin{figure*}
	\centering
        \includegraphics[width=0.49\linewidth]{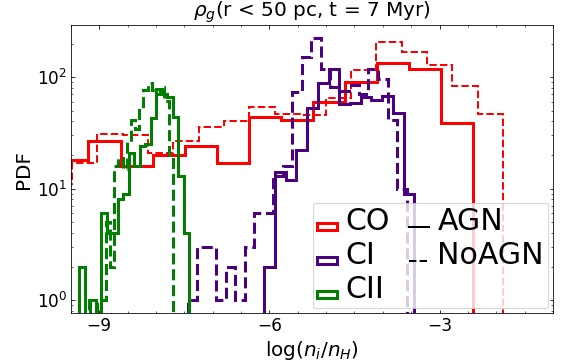}
        \includegraphics[width=0.49\linewidth]{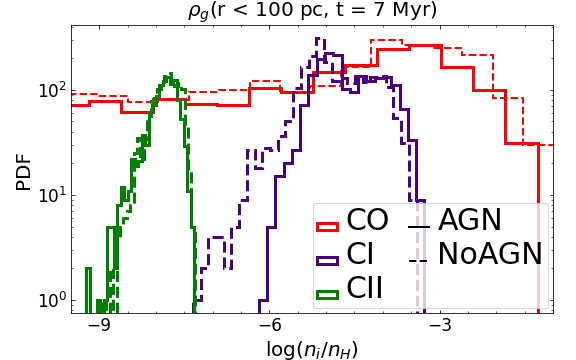}
	\caption{Probability Distribution Function (PDF) of number density ($n_i$) of CO(red), \text{CII} (green),and CI(purple) in unit of total hydrogen number density, $n_{H}$, for AGN(solid-lines) and NoAGN(dashed-lines) model in different radius of 50(left) and 100(right) pc scale of CND at $t = 7$ Myr  which is the timescale during which we can confirm the effects of the AGN. 
    }
\label{fig:PDF_ALL}
\end{figure*}

\cite{Saito2022} compares the CI/CO intensity ratios in different regions of NGC 1068, a nearby  starburst/AGN galaxy, and finds that they vary from 0.1 to 1.0, with higher values in the outflow regions and lower values in the disc regions. The study interprets these variations as the result of different excitation and optical depth effects, as well as different chemical abundances and destruction rates of CI and CO. They suggests that CI is more abundant and less affected by photodissociation than CO in the outflow regions, where the radiation field and the gas density are higher. They also suggests that CI is more optically thin than CO in the outflow regions, where the gas column density is lower.
A kinematic model indicated that the CI enhancement was predominantly driven by the interaction between the outflow and the disc, which compressed and heated the gas. This implies that the AGN feedback has a negative effect on star formation in the CND and several recent studies indicate that the CND of NGC 1068 is a region dominated by X-ray radiation \citep[XDR;][]{Usero2004,Perez-Beaupuits2009,Krips2008,Garcia-Burillo2010}. 

Another example is the nearby type 2 Seyfert galaxy, Circinus, which hosts one of the closest AGNs. 
\cite{Izumi2018} revealed that the CI (1-0)/CO (3-2) ratio at the AGN position was unusually high, suggesting an XDR-type chemistry \citep{Izumi2018,Miyamoto2018}.
This means that the AGN radiation can destroy molecules and create atomic gas, which can affect the star formation efficiency and rate \citep{Park2022}. 
The kinematics of both lines showed the multi-phase nature in the torus region, i.e., the diffuse atomic gas was more spatially extended along the vertical direction of the disc than the dense molecular gas. This was explained by a radiation-driven fountain model, which proposed that atomic outflows were driven by radiation pressure from the AGN and created a geometrically thick atomic disc. This supported the validity of the radiation-driven fountain scheme in explaining the physical origin of the AGN torus \citep{Uzuo2021}. 

CO-dark molecular gas, represents a significant component of the interstellar medium in galaxies, particularly in regions with intense radiation fields such as those near supermassive black holes in the central nuclear region (CNR) of galaxies \citep{MontoyaArroyave2023}. In Seyfert-2 type AGN-dominated galaxies, the presence of a supermassive black hole can influence the surrounding gas, potentially increasing the proportion of CO-dark gas. As shown by \cite{Madden2020}, this gas can be effectively traced by \text{CII} and CI emission lines, which highlight the reservoirs of molecular hydrogen (H$_2$) that CO fails to detect. The \text{CII} 158 micron line, in particular, is bright and can trace the total H$_2$ mass, revealing up to 70\% to 100\% of the H$_2$ mass that is not traced by CO in dwarf galaxies. Conversely, in galaxies without AGN, the star formation histories and the properties of the molecular gas can differ.

AGN feedback encompasses the various processes through which the active nucleus of a galaxy interacts with its surrounding gas \citep{Croton2006,Booth2009,Dubois2013,Bower2017,Raouf2017,Raouf2019,Raouf2024}, influencing its physical and chemical properties. In this study, our focus is on investigating the influence of AGN feedback on the transition from atomic to molecular gas within a CND surrounding a supermassive black hole (SMBH). Our investigation builds upon recent hydrodynamics simulation studies, HDGAS \citep[]{Raouf2023}, to further explore the effects of AGN feedback on the CND scale. The transition from neutral atomic to a molecular gas phase is especially interesting, as it reflects the balance between the various processes that form and destroy molecules, such as heating, cooling, shielding, and dissociation. The transition also affects the observability and detectability of the CND, as different gas phases emit or absorb different types of radiation, such as radio, infrared, or X-rays \citep{Izumi2018}.  

Our analysis concentrates on the atomic and molecular species encompassed within the CHIMES chemistry network, including CI, \text{CII} and CO. 
The structure of the paper is as follows: the simulation and methods are described in Section \ref{sec:simulation}. In Section \ref{sec:galaxie_template}, we introduce the observational galaxy templates used for comparison. Section \ref{sec:results1} presents our results and analysis, while Section \ref{subsec:sum} offers a summary and discussion.

\section{The \textsc{HDGAS} Simulations} \label{sec:simulation}
In our study, we utilize hydrodynamic simulations called HDGAS to explore the ISM within the CND of AGN-dominated galaxies, considering the influence of mechanical feedback from the AGN. These simulations incorporate the CHIMES non-equilibrium chemistry network to accurately represent radiative cooling and AGN heating. Focusing on Seyfert-2 type galaxies, such as NGC~1068, we develop a model of a gas disc surrounding a black hole. By comparing this model to one without AGN feedback, we investigate the effects of AGN feedback on the central region of the galactic disc. For a comprehensive understanding of the HDGAS hydrodynamic simulation model, we refer interested readers to \cite{Raouf2023} (referred to as Paper-I).

The simulations were run using Gizmo \citep{Hopkins2015}, a hydrodynamics and gravity code that implements several different hydrodynamics solvers. The initial conditions consist of gas-rich nuclear disc containing a black hole in its center with $n=10^6$ gas particles, where each particle initially has a mass of  $\sim100\ \msun$. This simulation is coupled with stellar feedback and ISM physics from FIRE-2 as described in \cite{Hopkins2018a}. The formation of stars is only possible in cold, molecular, and locally self-gravitating regions that have a number density above $\nh = 10^4 \, {\rm cm}^{-3}$ (as used in the most recent FIRE simulation studies e.g \citep{Torrey2020}).
With the CHIMES non-equilibrium chemistry and cooling model \citep{Richings2014a,Richings2014b}, chemical abundances, for 157 species  are calculated over time,  including all ionization states of 11 elements that are important for cooling, as well as for 20 molecules, including  CO, HCO$^+$ and  $\rm H_2O$. By integrating the temperature in time with the 157 rate equations, the CHIMES module calculates cooling and heating rates. 

We use the black hole accretion model implemented by \citet{Hopkins2011}. In this approach, BH accretion is assumed to be determined by the gravitational torques 
\citep[see][for more details of accretion methods]{Hopkins2016}. In this study, AGNs exhibit blue-shifted BAL or troughs when their line of sight is intercepted by a high-speed outflow, likely originated from their accretion discs \citep[which is unresolved in the simulation; see][]{Hopkins2011}.
Through the mechanical feedback process we implement in this simulation, wind mass and kinetic luminosity are continuously “injected” into the gas surrounding the SMBH where the outflow is isotropic \citep{Hopkins2016}.
It is assumed that some fraction of the photon momentum drives a wind at the resolution scale around the BH \citep{Murray1995}. Hence, the accreted gas is blown out as a wind with velocity $v_{\rm wind}$. The wind is defined by two parameters, the mass-loading of $\beta\equiv\dot{M}_{\rm wind}/\dot{M}_{\rm BH}$ and the velocity $v_{\rm wind}$ that equivalently relate to momentum-loading ($\dot{p}_{\rm wind}=\eta_{p}\,L/c$) and energy-loading ($\dot{E}_{\rm wind}=\eta_{E}\,L$) of the wind. 
In this study our simulation has been carried out using fixed energy and momentum loading factor for AGN model, \citep[with value for $\beta = 6$ and $v_{\rm wind}= 5000 km/s$ based on observations and theoretical models; see for e.g.][]{Moe2009,Dunn2010,Hamann2011,Borguet2013}. In the NoAGN model, we have set both $\beta$ and $v_{\rm wind}$ to zero.

The maps of emission lines are generated through the post-processing of simulation snapshots using version 0.40 of the publicly available Monte Carlo radiative transfer code  \textsc{radmc-3d} \citep{Dullemond2012}, utilizing the abundances of ions and molecules calculated during the simulations with the CHIMES chemistry module.
All line images are produced with a passband ranging from -100 km/s to +100 km/s and 400 wavelength points evenly distributed around the line's center. In this study, we generated emission maps for CI, CII, and CO. Specifically, we utilized the first J-line of atomic carbon (CI) at a frequency of 492 GHz, which is crucial for probing the warm, dense gas regions typical of AGN environments. We also examined ionized carbon (CII) at 1900 GHz, a key fine-structure line in the far-infrared spectrum. Additionally, we analyzed carbon monoxide (CO(2-1)) at 230 GHz, which is commonly used to trace molecular gas, essential for star formation. Using NGC 1068 as an example, we use the distance of 14 Mpc (so that 70 pc equals 1" in each moment map) and inclination of 41 degree. 

\begin{figure}
	\centering
        \includegraphics[width=0.99\linewidth]{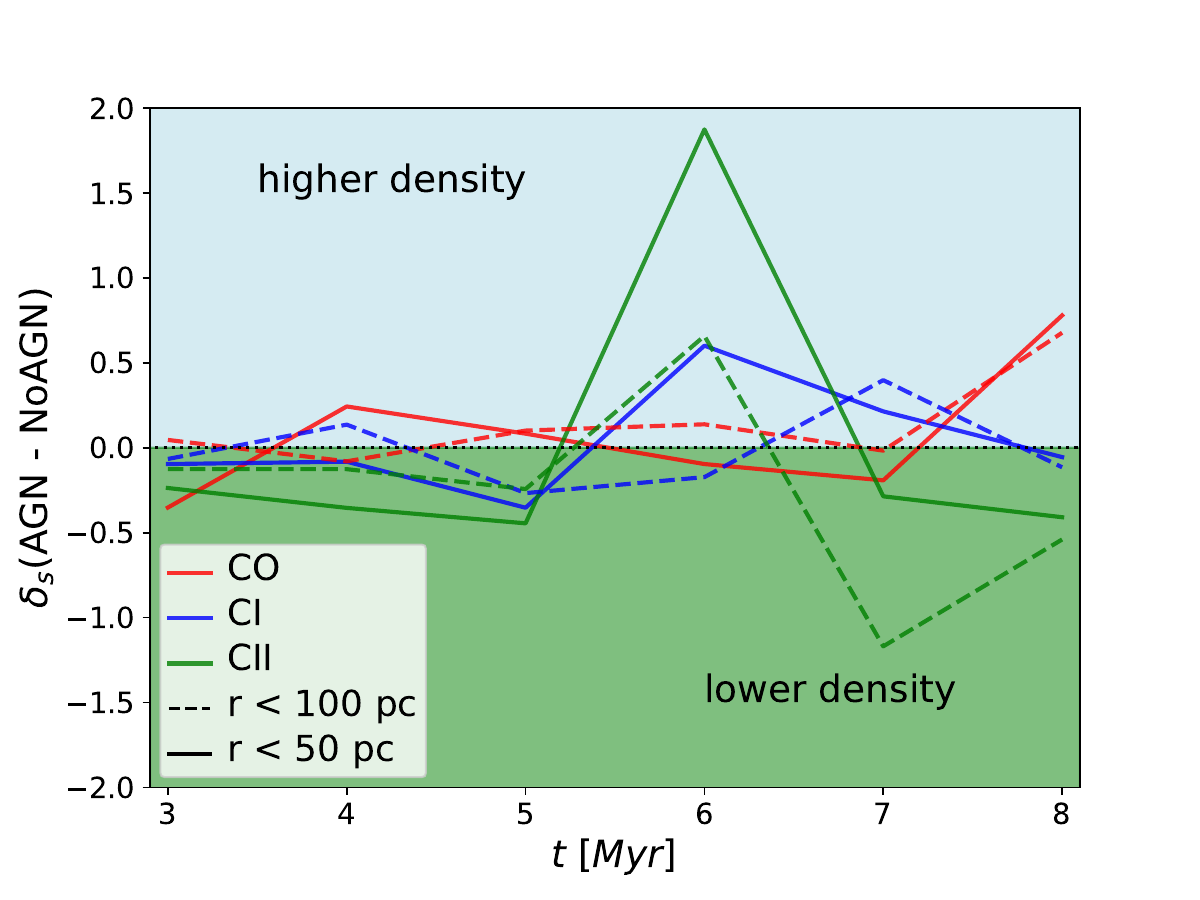}
	\caption{  Asymmetry ($\delta_s$, Skewness) offset in the PDF of the AGN and NoAGN models for CO (red), CI (blue), and CII (green) at different radii of 50 (solid lines) and 100 (dashed lines) pc in the CND. The asymmetry (Skewness) of the distributions highlights the differences between AGN and NoAGN data across various timescales. A positive value indicates $\delta_s$ (AGN) > $\delta_s$ (NoAGN), reflecting a longer tail on the right side and a higher number density tendency for AGN. Conversely, negative values ($\delta_s$ (AGN) < $\delta_s$ (NoAGN)) indicate a longer tail on the left side and a lower number density distribution for the AGN model.
    }
\label{fig:Skew_ALL}
\end{figure}

\begin{figure*}
	\centering
        \includegraphics[width=0.49\linewidth]{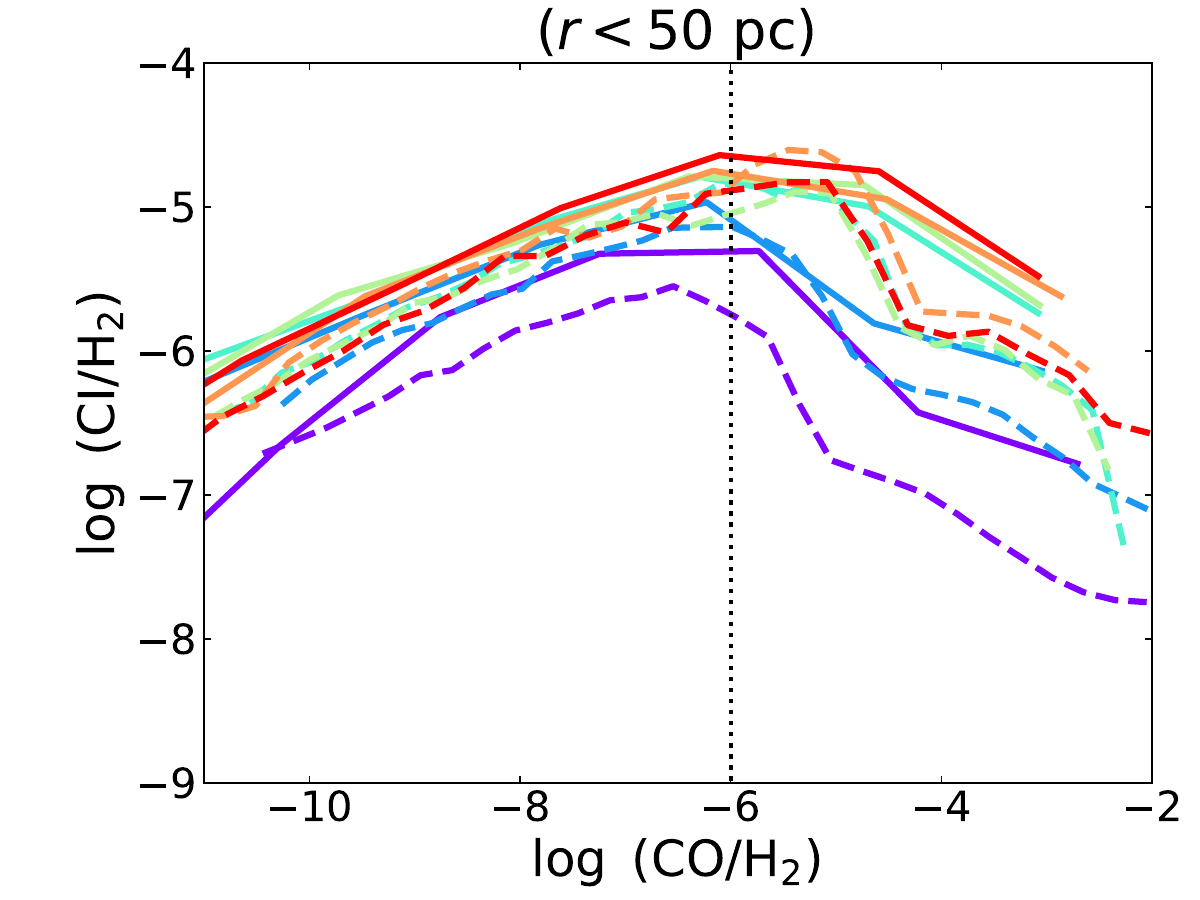}
        \includegraphics[width=0.49\linewidth]{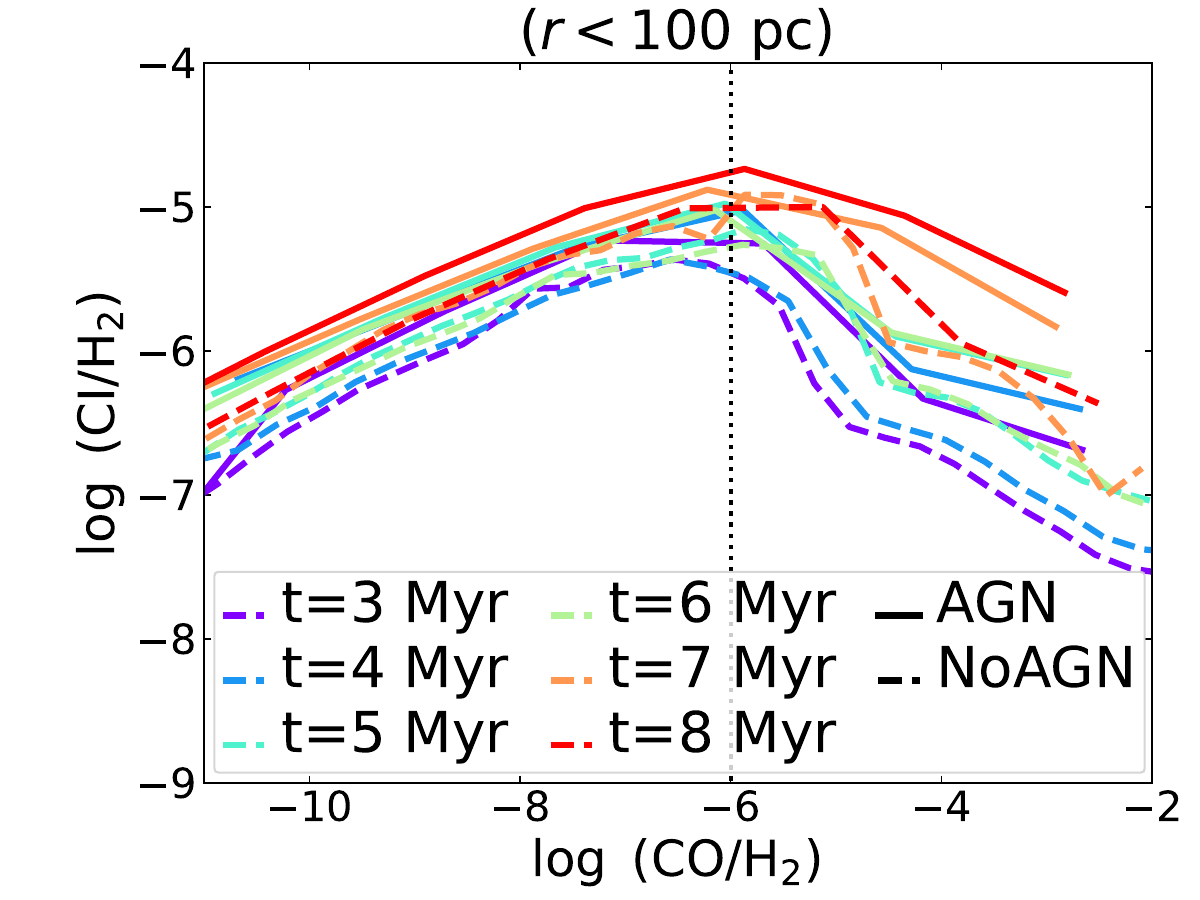}
        \includegraphics[width=0.49\linewidth]{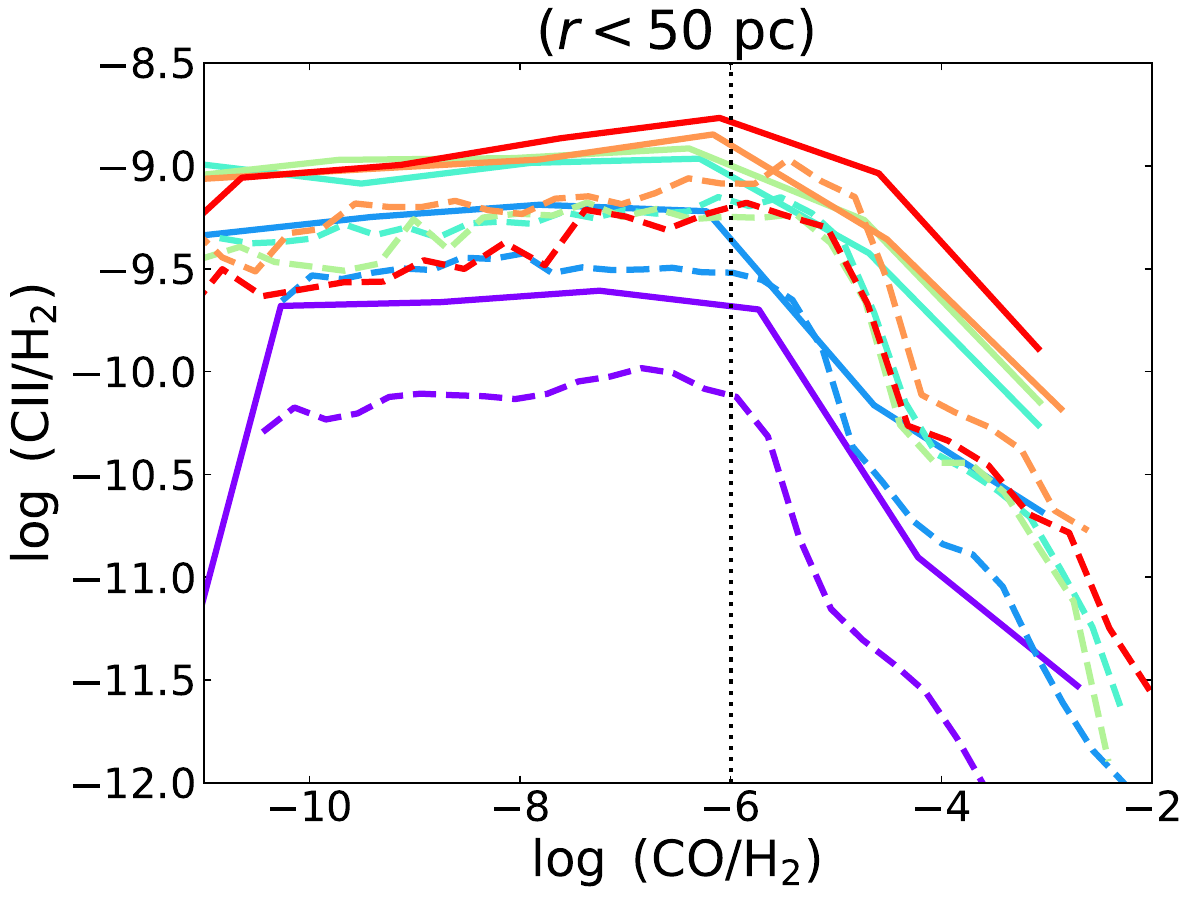}
        \includegraphics[width=0.49\linewidth]{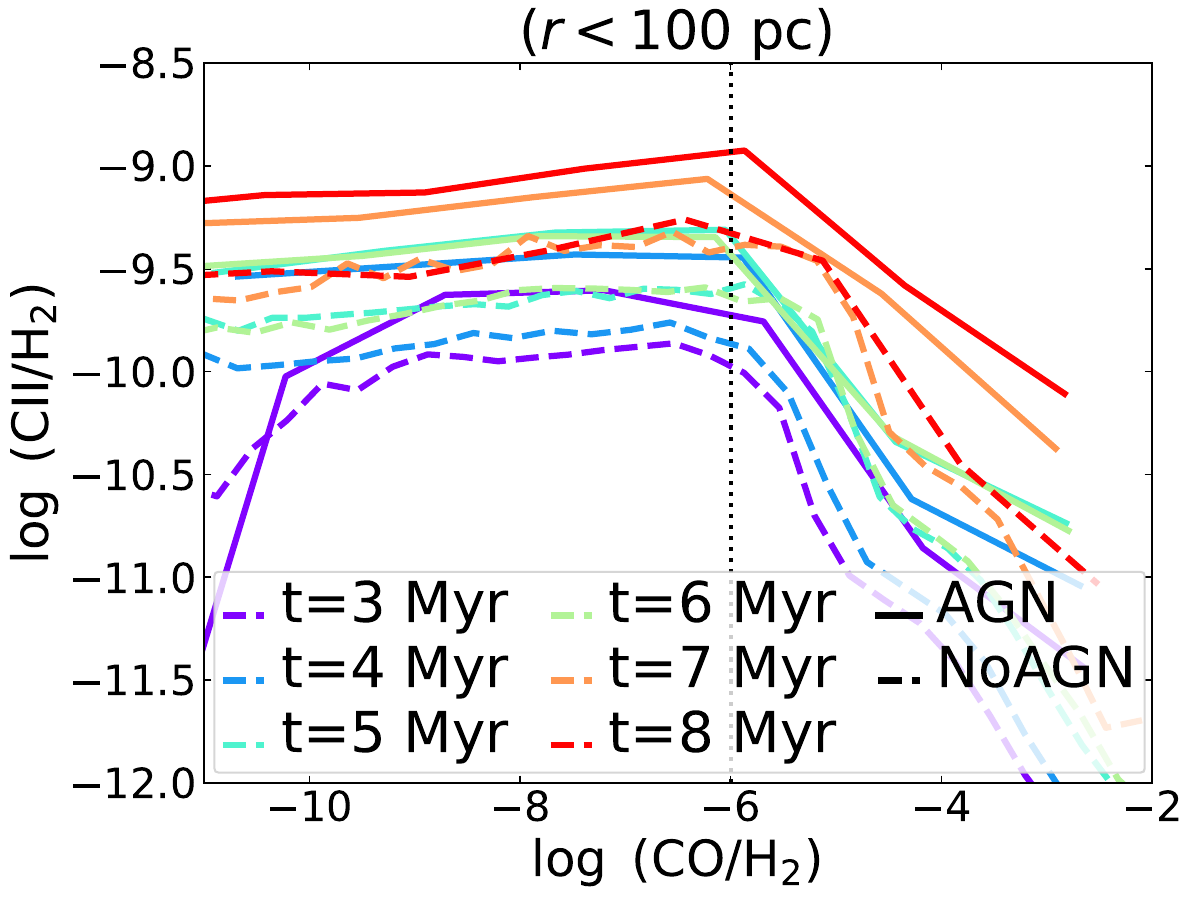}   
	\caption{ Abundance distribution of CI (top) and CII (bottom) as a function of CO density normalized to $H_2$ for AGN and NoAGN models at disc radii of < 50 pc (left) and < 100 pc (right). The top panels show a correlation between CI/H$_2$ and CO/H$_2$ ratios at low CO abundances (CO/H$_2 < 10^{-6}$), with an anti-correlation at higher ratios ($> 10^{-6}$). The AGN model exhibits higher CI/H$_2$ ratios, particularly within 50 pc, indicating enhanced CI abundance due to AGN feedback. The bottom panel reveals a steeper anti-correlation for CII/H$_2$ at high CO/H$_2$ ratios, with the AGN model showing consistently higher ionized CII abundances. The vertical dotted line indicates CO/H$_2 = 10^{-6}$, marking shifts in correlation.
    }
\label{fig:scatt_CI_CO}
\end{figure*}

\begin{figure}
	\centering
 \includegraphics[width=1.05\linewidth]{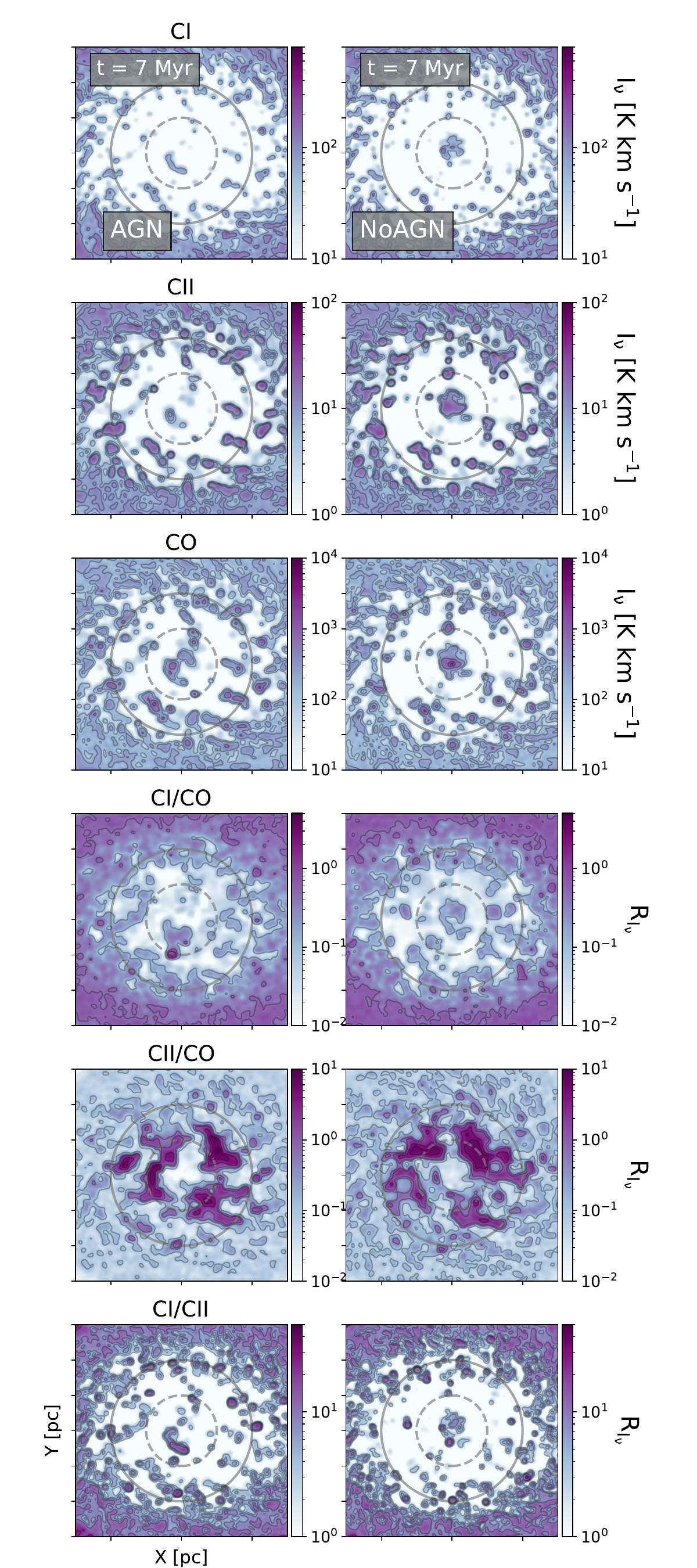}
	\caption{The first three panels showcase integrated intensity maps (Moment-0) for CI (top) and CO (middle) lines, along with their ratio CI/CO (bottom), generated using the RADMC-3D radiative transfer code applied to simulation data for the first J-line emission (J=1-0) at 7 Myr for AGN (left) and NoAGN (right) model. A consistent beam size of 0.06" x 0.06" is applied throughout the images, corresponding to a physical scale of 4.2 parsecs at the comoving radial distance of 14 Mpc for NGC 1068, with a position angle of zero degrees. Each panel depicts a central 150 pc region of the disc. Over-intensity regions in each map are highlighted by contours within the range of their respective color bars. Concentric circles represent the central 50 and 100 pc radii of the disc.
 }
	\label{fig:M0_CI_CO}
\end{figure}

 \begin{figure*}
	\centering
 \includegraphics[width=0.33\linewidth]{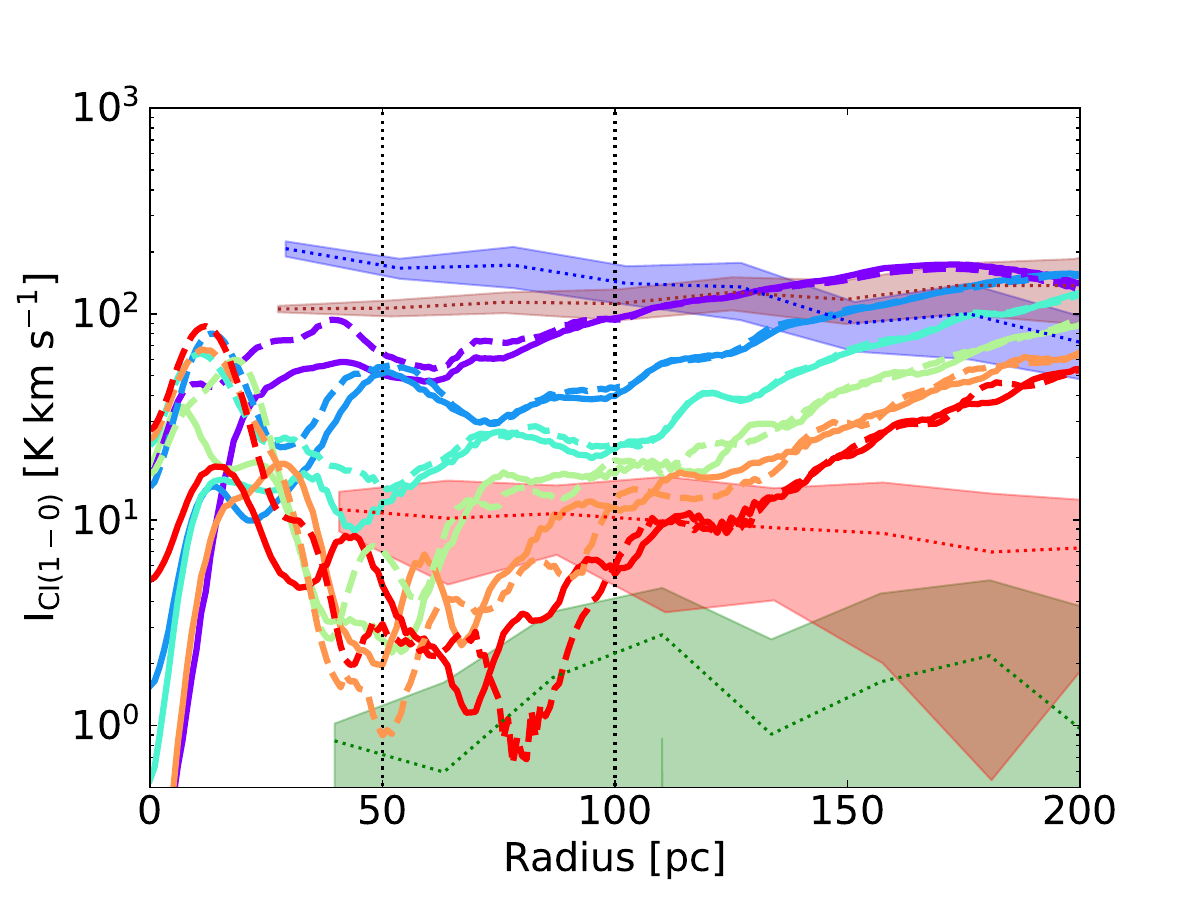}
  \includegraphics[width=0.33\linewidth]{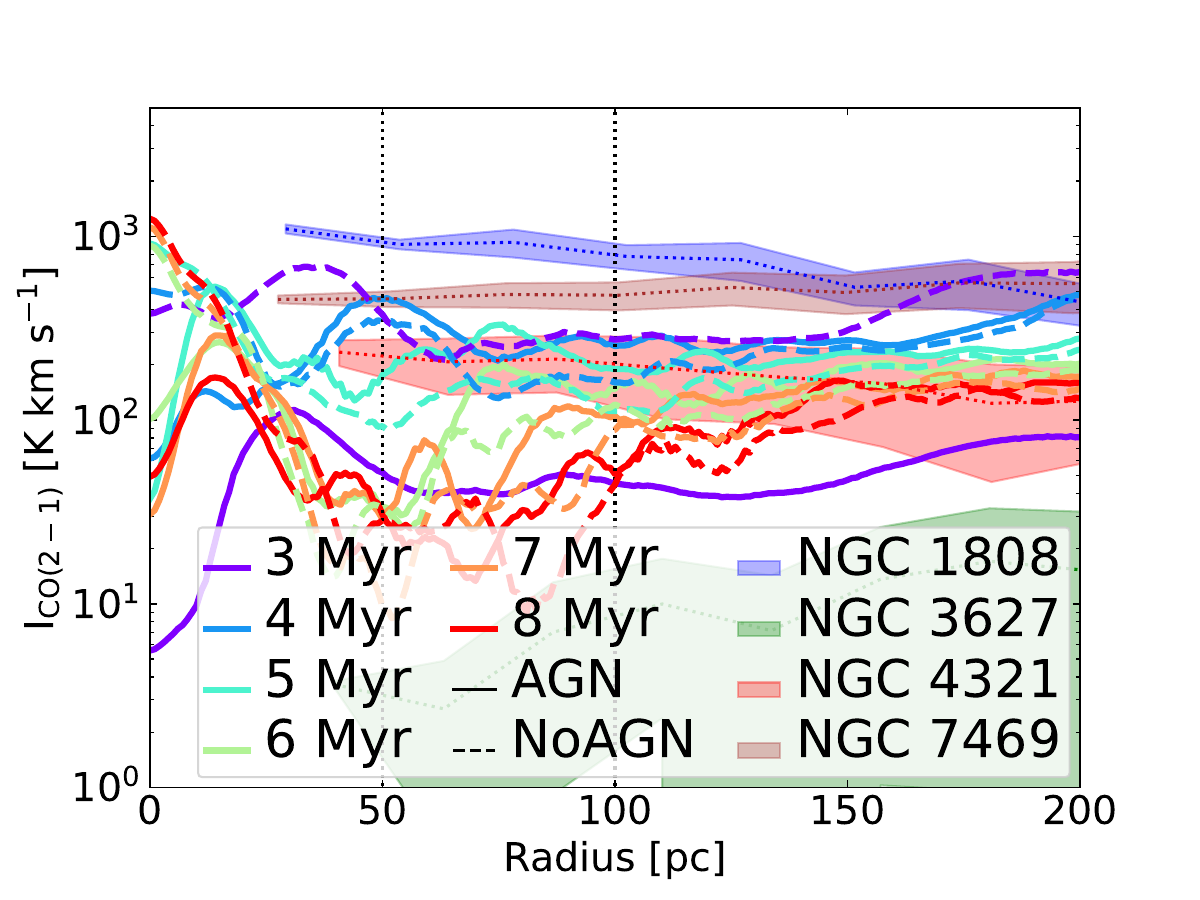}
 \includegraphics[width=0.33\linewidth]{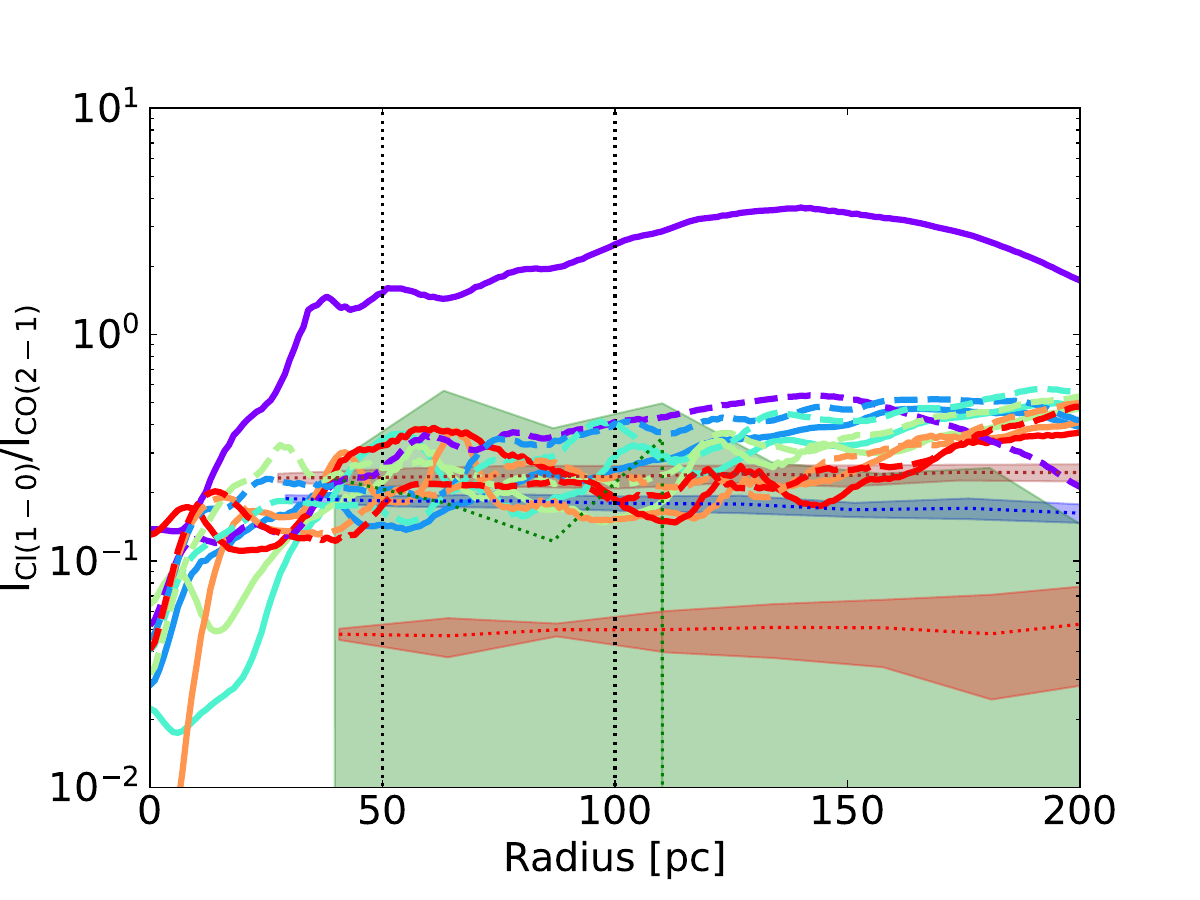}
    \includegraphics[width=0.33\linewidth]{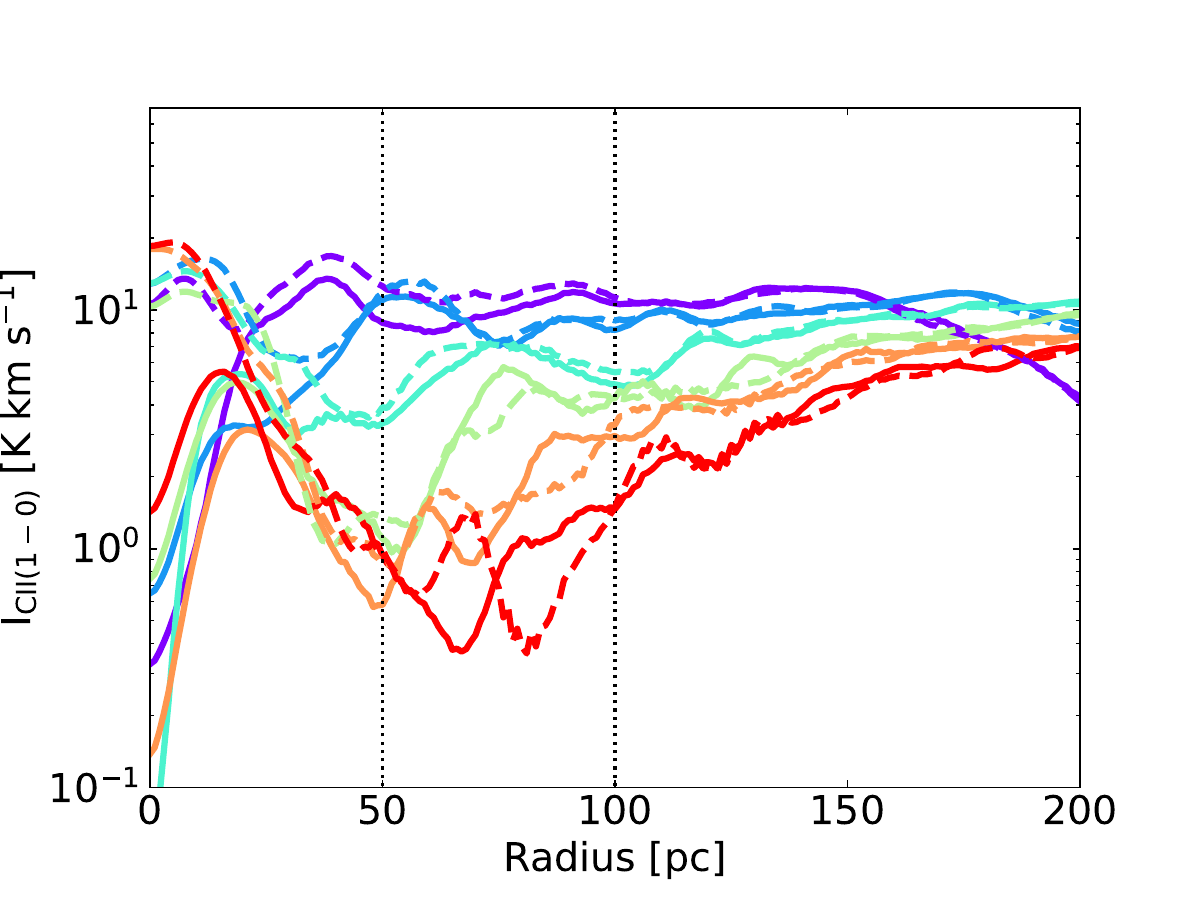}
   \includegraphics[width=0.33\linewidth]{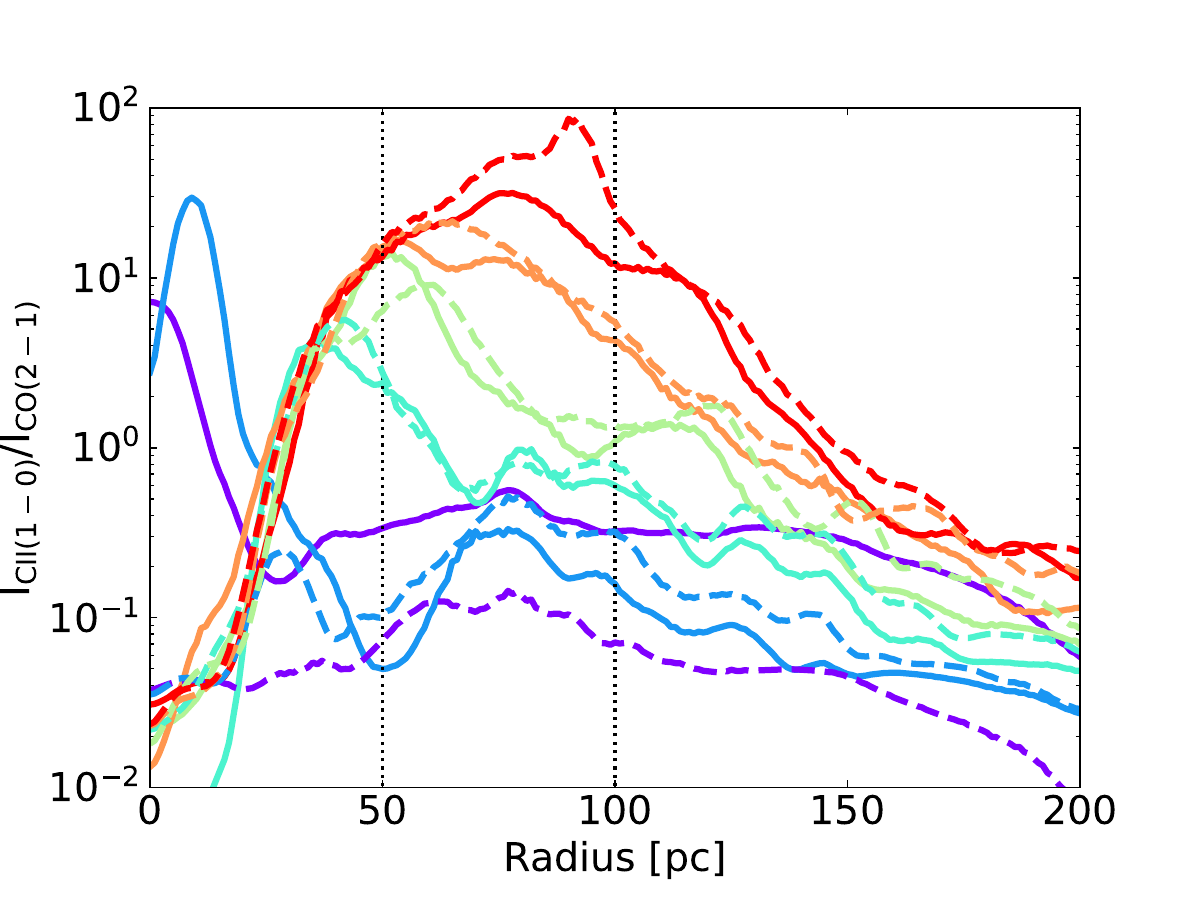}
 \includegraphics[width=0.33\linewidth]{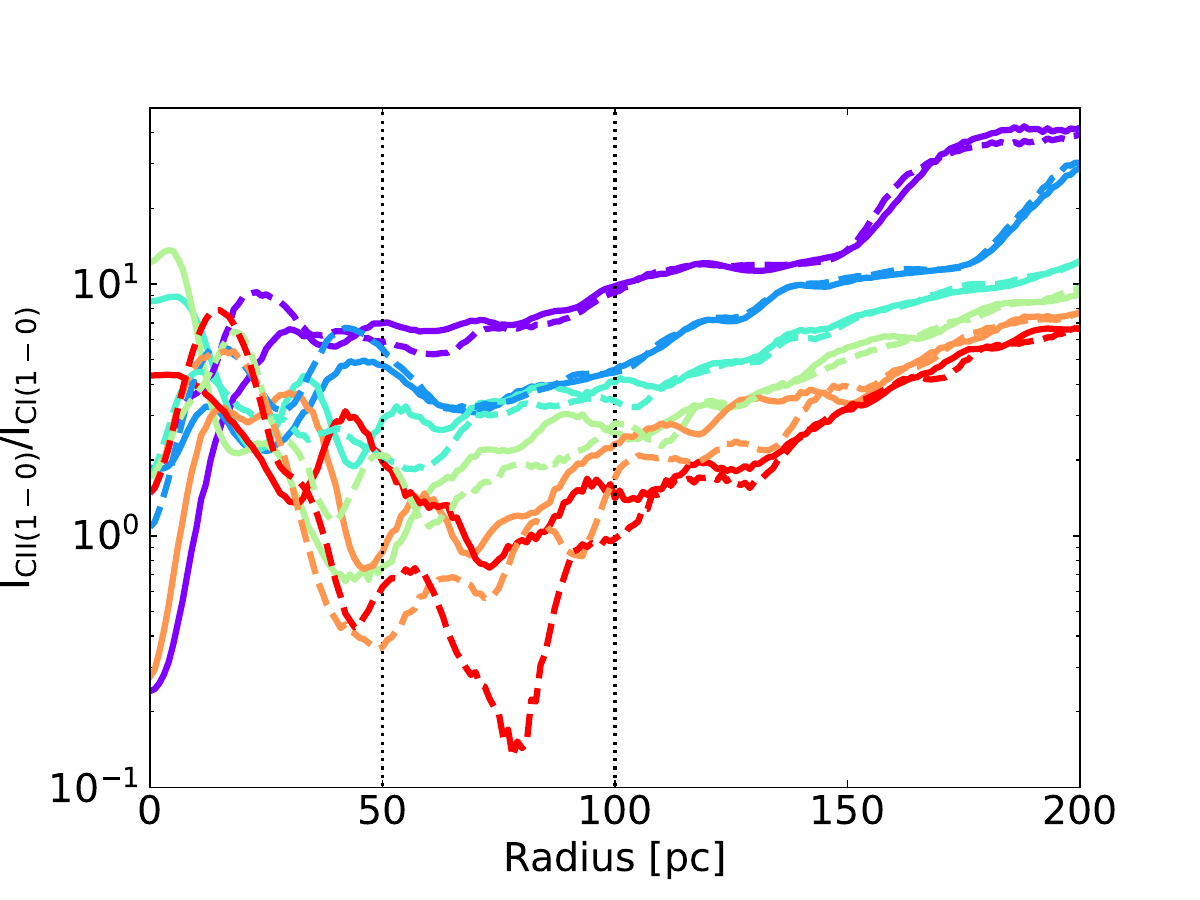}
	\caption{
     Top: Radial distribution profiles of CI (left panel), CO (middle panel), and CI/CO ratio (right panel) intensity maps for AGN (solid lines) and NoAGN (dashed lines) models over a time range of 3 to 8 Myrs. The NoAGN model shows higher intensity values below 50 pc. The CI/CO ratio in the NoAGN model is significantly higher at radii below 50 pc (t > 3 Myr), indicating a stronger interface of ionized and molecular gas without AGN influence. At t = 3 Myr, the AGN model displays a higher CI/CO(2-1) ratio across all radii, and at t = 8 Myr, it shows increased CI/CO ratios between 30 and 100 pc, suggesting AGN effects on gas dynamics. The color-shaded lines represent observed trends for galaxies with (e.g., NGC 1808) and without AGN (e.g., NGC 3627). 
    Bottom: Similar profiles for CII (left panel), CII/CO ratio (middle panel), and CI/CII (right panel) intensity maps. The CII emission is higher in the NoAGN model's central region, while both models show a decrease in CII emission in outer regions beyond 50 pc over time. The AGN model initially has a higher CII/CO(2-1) ratio in central regions, but this reverses in outer regions at later times, indicating a decrease in relative CII to CO emission. The CI/CII ratio profile reveals that the AGN model shows a lower ratio in the central region initially, which reverses at larger radii (beyond 30 pc) at later times, suggesting changing relative emissions as the galaxy evolves.
 }
	\label{fig:Profile_CI_CO}
\end{figure*}

 \begin{figure*}
	\centering
\includegraphics[width=0.49\linewidth]{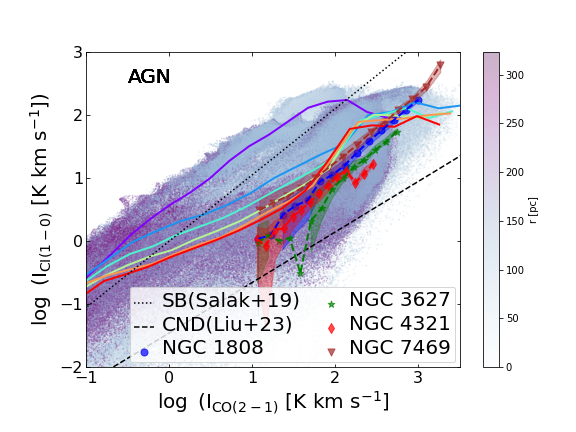}  
\includegraphics[width=0.49\linewidth]{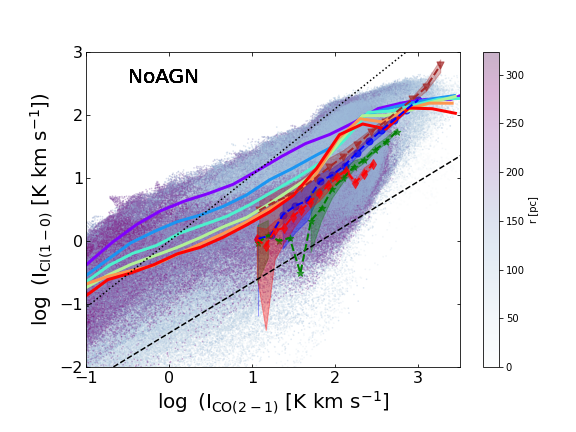}
\includegraphics[width=0.49\linewidth]{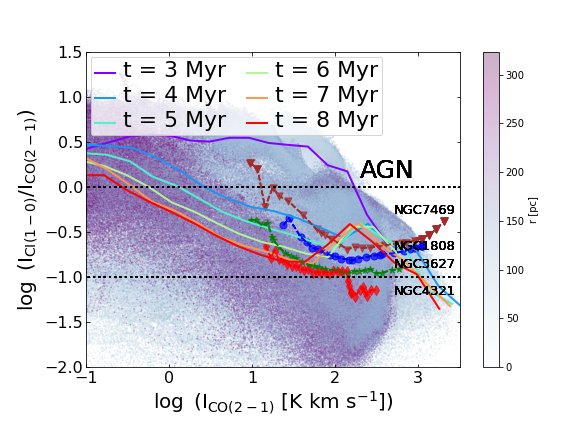}   
\includegraphics[width=0.49\linewidth]{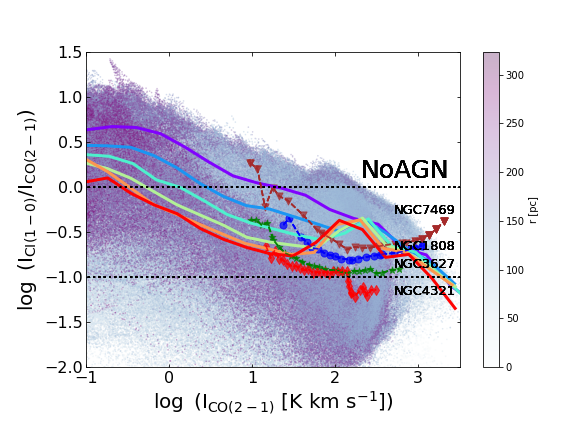}
\caption{ 
 Scatter plot of CI brightness ($I_{\rm CI}$, top) and CI/CO ratios ($I_{\text{CI}}/I_{\text{CO}}$, bottom) versus CO brightness ($I_{\text{CO}}$) across different time scales (3-8 Myr) and disc radii for AGN (left) and NoAGN (right) models. Points are color-coded by distance from the center of the disc. A correlation is observed between CI brightness and CO, consistent with trends in starburst regions \citep[dotted line,][]{Salak2019} and circumnuclear discs \citep[dashed line,][]{Liu2023}, with AGN models exhibiting higher CI brightness. 
The bottom panels illustrate that simulations predict higher CI/CO ratios at fainter CO pixels, particularly at outer radii, contrasting with observations where upward tails in CI vs. CO at the CO-faint end are primarily due to detection limits. The simulations accurately capture this behavior, showing higher CI and CI/CO values for AGN models even at early times (above the dashed line representing the identity value \( \log (I_{\text{CI}}/I_{\text{CO}}) > 0 \)). Systematic differences in CI/CO ratios are noted among galaxies, with NGC 3627 and NGC 4321 displaying a tight distribution around 0.1 \citep[dashed line,][presenting star forming disk]{Liu2023}, while NGC 1808 and NGC 7469 show increased ratios in starburst regions, reaching about 0.2, indicating different galactic environments. The color bar indicates the distance from the disc center, and shading lines in the top panels represent the $\sigma$ levels for each part of the CI map within 200 pc.
}
	\label{fig:intensity_scatter_CO_CI}
\end{figure*}

\section{Galaxies templates}\label{sec:galaxie_template}

We have used submmilimeter ALMA observations  of a selection of galaxies to compare our simulated maps. We specifically focused on four galaxies\footnote{We selected NGC 1808, NGC 7469, NGC 3627, and NGC 4321 for comparison because these galaxies have been observed in terms of CI and CO with the same resolution, allowing for a more consistent and accessible comparison between AGN and NoAGN cases. This uniformity in observational data is crucial for our analysis. Additionally, NGC 1068 is being utilized in our other studies, where we focus on its higher resolution observations. Thus, we aimed to maintain a clear distinction between the datasets used for different analyses while ensuring that the selected galaxies provide reliable comparisons within the context of our current work.
}: NGC1808, NGC7469, NGC3627, and NGC4321. Note that all the selected galaxies are nearby and comparable in distance to NGC 1068.

NGC1808 is a galaxy exhibiting intense starburst activity. 
Observations of CI(1-0) and CO(2-1) have revealed a high column density of atomic carbon in the circumnuclear disc \citep{Salak2019}. Furthermore, the CI/CO intensity ratios suggest that the luminosity of CI can serve as a reliable tracer of molecular gas mass, particularly in resolved starburst nuclei \citep{Salak2019}.
In the case of NGC7469, a Seyfert 1 galaxy, the circumnuclear gas disc (CND) exhibits a ring-like structure and a two-arm/bi-symmetric spiral pattern. Surrounding the CND is a starbursting ring. By studying the emissions of CI (1-0) and CO (1-0), 
\cite{Nguyen2021} measures the mass of the supermassive black home (SMBH) via the CI emission which seems to be a better mass probe than the CO.
NGC3627, also known as M66, is a galaxy that forms part of the Leo Triplet.  ALMA observations \citep{Liu2023} have shown nearly uniform CI/CO line ratios across the majority of its star-forming disc. There is also  an excellent spatial correspondence between the CI and CO emissions. The abundance ratio of [CI/CO] in NGC3627 has been estimated to be approximately 0.1, consistent with previous large-scale studies conducted on the Milky Way \citep{Liu2023}.
Similarly, observations of NGC4321 also referred to as M100, within its inner discs have revealed nearly uniform line ratios of CI and CO emissions across its star-forming discs. \cite{Liu2023}  have observed a mild decrease in the CI/CO ratio with increasing metallicity. The abundance ratio of [CI/CO] in NGC4321 has also been estimated to be around 0.1, indicating a consistent chemistry with NGC3627.

Note that, these observations generally have a lower resolution (with an average spatial resolution greater than 100 pc.) than our simulations, which may affect our interpretations. However, our aim is to highlight the overall trend of our results in relation to the observations, rather than to focus on the specific details of comparison.

\section{Analysis and Results} \label{sec:results1}

\subsection{Atomic and Molecular gas Abundances}

In this study, we analyze the distributions and probability density functions (PDFs) of CI, CII, and CO for different regions of the circumnuclear disc (CND) with radii lower than 50 and 100 pc.  The results are shown in Figure \ref{fig:PDF_ALL} for 
$t = 7$ Myr, which is the timescale during which we can confirm the effects of the AGN. This snapshot is optimal since the gas depletion timescale is under 1 Myr \citep{Garcia-Burillo2014}, making a simulation duration of 10 Myr ideal. By normalizing the PDFs to the total hydrogen number density ($n_{H}$), we can directly compare the variations in molecular and atomic gas densities, providing an understanding of their respective contributions and the overall gas distribution within the system. Although the density peaks in the PDFs do not show significant differences between the AGN and NoAGN models, the skewness and tails in the high and low-density regions are crucial indicators of AGN influences.
 In the region \( r < 50 \, \text{pc} \), the peak of the logarithm of CO abundance relative to hydrogen number density \( \left( \frac{n_{\text{CO}}}{n_{\text{H}}} \right) \) is \(-3.7\) for both the AGN and NoAGN models. This similarity suggests that the conditions leading to CO formation are comparable in both scenarios, despite the presence of an active galactic nucleus (AGN) in one model. However, as we look at the carbon species, we see a divergence: for CI, the logarithmic abundance peaks at \(-5.5\) for NoAGN, compared to \(-5\) for AGN. This indicates that the presence of an AGN potentially enhances the CI abundance, possibly due to increased heating or photodissociation processes that favor CI formation. For CII, the peaks are even lower, with values of \(-8.3\) for NoAGN and \(-8\) for AGN. This further highlights the influence of the AGN, which likely provides additional energy that facilitates the ionization processes necessary for CII formation. The differences in CI and CII suggest that the AGN environment promotes a higher degree of ionization, affecting the chemical pathways of carbon species. In the region \( r < 100 \, \text{pc} \), the CO probability density function (PDF) peaks at \(-4\) for NoAGN, while the PDF peak is at \(-3.5\) for the AGN model. This shift indicates that the AGN environment may contribute to a higher CO abundance, perhaps through shock heating or enhanced star formation activity associated with the AGN. Furthermore, the peaks of the PDF for CI in both models are \(-5.3\) for NoAGN and \(-5\) for AGN, reinforcing the idea that the AGN fosters conditions that are more conducive to CI formation. The differences in PDF peaks reflect the varying chemical environments influenced by the AGN presence.
Finally, the peaks of the PDF in CII for both the AGN and NoAGN models are \(-7.9\), suggesting that while the AGN impacts the other carbon species, the CII formation might reach a saturation point in the conditions present, resulting in similar CII levels across both models.

Furthemore, we examine the skewness\footnote{Incorporating the PANDAS Python package \citep{reback2020pandas}  ($\delta_s$) for estimations of skewness.} of the distributions in Figure \ref{fig:Skew_ALL} to assess the asymmetry between the AGN and NoAGN models for time scales between 3 to 8 Mrys. A positive skewness value ($\delta_s > 0$) indicates a longer tail on the right side of the distribution, suggesting higher density regions. Conversely, a negative skewness value ($\delta_s < 0$) corresponds to a longer tail on the left side, representing lower density regions.

Figure \ref{fig:PDF_ALL} illustrates a bimodal distribution of CI density (purple lines) in the PDFs for both the AGN and NoAGN models across various radii. 
The density exhibits fluctuations in low-density areas (early time; t < 5 Myr) and high-density areas (late time; t > 5 Myr), suggesting an increase in CI density over time for the AGN model (see Figure \ref{fig:Skew_ALL}; blue lines). 
 In the CO skewness ($\delta_s(\text{AGN-NoAGN})$), the temporal trend indicates higher density tails for the AGN model at both 50 pc and 100 pc scales of the disc. Specifically, at t = 7 Myr (shown in Figure \ref{fig:PDF_ALL}), the AGN model is indeed located in lower density regions compared to the NoAGN model. However, we also emphasize that the overall trend from t=3 to t=8 Myr reveals distinct behaviors for the AGN and NoAGN models, especially in regions greater than 50 pc. This distinction highlights the influence of AGN feedback on the gas distribution. 
 For CII, the deviations in the PDFs are notably smaller compared to those for CI. As illustrated in Figure \ref{fig:Skew_ALL}, CII predominantly exists in the lower density mode. The AGN model shows a significant change in skewness, transitioning from \( \delta_s(\text{AGN}, t=5 \text{ Myr}) = -1.54 \) to \( \delta_s(\text{AGN}, t=6 \text{ Myr}) = 0.84 \). In contrast, the NoAGN model exhibits a smaller change, with skewness values of \( \delta_s(\text{NoAGN}, t=5 \text{ Myr}) = -1.10 \) and \( \delta_s(\text{NoAGN}, t=6 \text{ Myr}) = -1.03 \). This shift in skewness for the AGN model indicates a transition to higher density regions at 6 Myr, particularly in the inner regions ($<$ 50 pc).

Furthermore, we investigate the correlations between CI, CII, and CO abundances normalized to $H_2$ within the CND. The results are visualized in Figure \ref{fig:scatt_CI_CO}, shedding light on the distinct effects of AGN feedback on the molecular composition of CO at different radii. In the top panels of Figure \ref{fig:scatt_CI_CO}, we show the CI/H$_2$ ratio against the CO/H$_2$ ratio. The data points exhibit a correlation at low CO abundances (CO/H$_2< 10^{-6}$). However, an intriguing anti-correlation trend (between CI and CO) emerges at high CO/H$_2$ ratios ($> 10^{-6}$) for both the AGN and NoAGN models. Further, the AGN model displays significantly higher CI/H$_2$ ratios compared to the NoAGN model, particularly within radii less than 50 pc. This suggests that AGN feedback is particularly effective in enhancing the abundance of CI in the inner regions of the disc, likely due to increased radiation fields, heating of the interstellar medium, gas mixing from outflows, and the formation of extensive photo-dissociation regions. 
In the bottom panel of Figure \ref{fig:scatt_CI_CO}, we examine the CII/H$_2$ ratio as a function of the CO/H$_2$ ratio. We see a steeper anti-correlation for CO/H$_2> 10^{-6}$ compared to CI/H$_2$. Also, we see that at lower CO abundance levels, the abundance of CII remains relatively constant and the AGN model exhibits higher levels of ionized CII gas abundances at different time scales (most of the times). The intense photoionization or dissociation processes induced by the AGN likely contribute to this effect.

 Moreover, the higher CO/H$_2$ ratios observed in the NoAGN model suggest that the AGN is influencing these ratios, particularly in both regions beyond 100 pc and 50 pc from the disc, where we do not observe $\log(\text{CO/H}_2)$ values exceeding approximately -3 in the AGN model. This indicates that the presence of the AGN leads to a suppression of the CO/H$_2$ ratio in these areas, causing the solid lines to terminate at lower CO/H$_2$ values compared to the dashed lines.

\subsection{Integrated Intensity Maps and Profiles} \label{sec:profile}

Figure \ref{fig:M0_CI_CO} shows the spatial distribution of CI, CII, and CO integrated intensities, as well as their ratios (CI/CO, CII/CO, and CI/CII) in the circumnuclear disc for both the AGN and NoAGN models. 
The integrated intensity maps can reveal structures like molecular spiral arms, rings, or clumps, which may be associated with ongoing star formation or other phenomena related to the interaction between the AGN and the surrounding gas. 
The CI/CO ratio maps (t = 7 Myr) generally illustrate the extent of CO photodissociation and/or CII recombination, highlighting their effects on the molecular gas abundance in the photodissociation regions (PDR). Higher CI/CO ratios indicate decreased CO molecule levels due to photodissociation by AGN radiation.

In the NoAGN model, the high-intensity regions within 50 pc correspond to areas with a greater concentration of molecular gas, likely linked to active star-forming regions or dense molecular clouds. These regions exhibit stronger CI emission due to the higher abundance of carbon-bearing molecules undergoing CO photodissociation. Conversely, the AGN model reveals low-intensity regions within the same radius, indicating lower molecular gas densities or reduced CI and CO abundance. These areas may be influenced by AGN radiation, shocks, or other physical processes that affect the excitation or destruction of atomic CI and CO molecules, highlighting that the factors responsible for CO destruction differ from those that ionize CI (see CII in 2nd panels of Figure \ref{fig:M0_CI_CO} ).
Figure \ref{fig:Profile_CI_CO} illustrates the radial profiles of CI and CO intensity for both the AGN and NoAGN models. The NoAGN model displays higher intensity values for radii below 50 pc compared to the AGN model.
In the lower panels of Figure \ref{fig:M0_CI_CO}, the CI/CO(2-1) ratio intensity maps reveal characteristics of the photon-dominated region (PDR). 
  In Figure \ref{fig:Profile_CI_CO}, the NoAGN model displays a higher CI/CO ratio at radii below 50 pc for t$>$3 Myr, suggesting a larger interface between neutral atomic gas and molecular gas in the absence of AGN influence. Furthermore, while the overall trend does indicate an increase in the CI/CO ratio as a function of radius, the AGN model exhibits a higher CI/CO(2-1) ratio at t=3 Myr across all radii compared to the NoAGN model and other time points. 

 At later times (\( t = 8 \) Myr), the AGN model shows a higher CI/CO ratio for radii between 30 and 100 pc, indicating that the presence of an AGN influences the gas dynamics in this region. A high CI/CO ratio reflects the balance between different carbon phases in the interstellar medium. Specifically, a higher ratio suggests a greater presence of CI compared to CO, which may arise from processes such as increased star formation or heating associated with AGN activity.
The color-shaded lines illustrate the observed trends of CO, CI, and CI/CO profiles for samples without AGN, specifically NGC 3627 and NGC 4321, as well as for AGN samples like NGC 1808 and NGC 7469, within the central 200 pc regions (see Section \ref{sec:galaxie_template}). 
The AGN samples show higher CI and CO intensities across all regions (limited by resolution) and exhibit a consistent trend in CI/CO ratios similar to simulations at different time scales beyond 50 pc, although they are generally higher compared to the without AGN samples.

The bottom panels of Figure \ref{fig:Profile_CI_CO} shows the radial profile of CII and CII/CO and CI/CII integrated intensity maps illustrate in the bottom of Figure \ref{fig:M0_CI_CO}. The CII profile shows distinct differences between the AGN and NoAGN models. For the NoAGN model, the CII emission is higher in the central region across all timescales.
However, in the outer regions beyond $\sim$50 pc, the CII profiles for both models drop off inversely at later time steps, indicating a decrease in CII emission in the outer galaxy over time.
The CII/CO(2-1) ratio profile also exhibits interesting trends. At earlier time steps, the AGN model has a higher CII/CO(2-1) ratio, especially in the central regions. But at later time steps, this ratio increase in the outer parts (beyond $\sim$50 pc) for both the AGN and NoAGN models (t = 7,8 Myr), suggesting a decrease in the relative CII to CO(2-1) emission in the outer galaxy over time, regardless of the presence of an AGN. 

 Lastly, the CII/CI ratio profile reveals that in the earlier time steps, the AGN model has a lower CII/CI ratio compared to the NoAGN model, particularly in the central regions. This trend reverses at larger radii (beyond \(\sim 30\) pc), where the AGN model exhibits a higher CII/CI ratio than the NoAGN model at later time steps (\( t = 7, 8 \) Myr). This indicates that the relative amount of CII to CI emission is initially lower in the AGN model's central region, but becomes higher in the outer parts of the galaxy at later times.

\subsection{The Brightnesses in Star-Forming and Starburst disc Environments} \label{sec:environments}

In Figure \ref{fig:intensity_scatter_CO_CI}, we present scatter points representing the observed brightnesses of CI and CO, along with the CI/CO line ratios color coded by distance to the center of disc ($r$). A correlation is revealed for differert time scales (t = 3-8 Myr) between CI brightness as a function of CO, aligning with the range of observations from starburst regions \citep[SB]{Salak2019} and the circumnuclear discs (CND) \citep{Liu2023}. Notably, CI brightness is higher in AGN models, which is consistent with trends observed in galaxies hosting AGN.  In general, the observation trend is confined within the range of \( 1 < \log(I_{\text{CO}}) < 3 \). However, the simulation shows an extension to lower CO values in the range \( -1 < \log(I_{\text{CO}}) < 3 \) for both the AGN and NoAGN models. The scatter of CI intensity brightness is approximately \(-2 \, \text{dex}\), while the CI/CO ratio exhibits a scatter of around \(-2.5 \, \text{dex}\). The lower CI/CO ratios are found at greater distances from the center (\( r > 200 \, \text{kpc} \)). Time evolution leads to a lower CI/CO ratio at larger distances from the disc center (\( r > 200 \, \text{kpc} \)) for both the AGN and NoAGN models.

 As shown in the bottom panels of Figure \ref{fig:intensity_scatter_CO_CI}, the simulations predict higher CI/CO ratios at fainter CO pixels, especially in the outer regions. However, it is important to note that the outer regions of the simulations actually display lower ratios than the observations. The upward tails observed at the CO-faint end in the CI versus CO plot are primarily a result of the observed resolution limits of each galaxy. In contrast, the simulations provide a more accurate representation of this behavior, reflecting the underlying gas dynamics without the same detection constraints. Additionally, even at earlier time steps, the simulations indicate that both CI and CI/CO ratios are higher in the AGN model compared to the NoAGN model. Overall, the simulations show both higher and lower CI/CO ratios at fainter CO pixels, contributing to the scatter observed in the data.

 Additionally, systematic differences in CI/CO ratios are observed among these galaxies and within their galactic environments. The majority of molecular gas in the 3–7 kpc star-forming discs of NGC 3627 and NGC 4321 exhibits a distribution of CI/CO ratios centered around 0.1 (indicated by the dotted line in the bottom panels of Figure \ref{fig:intensity_scatter_CO_CI}) \citep{Liu2023}. It is important to note that this apparent tightness in the distribution, as shown in Figure \ref{fig:intensity_scatter_CO_CI}, results from the binning process applied to the original data points, which helps to emphasize the underlying trends while smoothing out individual variations.

This distribution remains relatively flat over an order of magnitude in CO(2-1), which is referred to as the star-forming discs regime (as described in Figure 5 \citep{Liu2023}). In contrast, the CI/CO ratio increases to about 0.2 within a galacto-centric radius of a few hundred parsecs to approximately 1 kpc in the starburst regions of NGC 1808 and NGC 7469, marking the starburst disc regime, as indicated by the dotted line at the identity value of \( I_{CI} \) and \( I_{CO} \).

\begin{table*}
    \centering
    \caption{  Summary of trends in median intensities of CI, CII, CO, and their ratios in AGN and NoAGN models, categorized by inner regions (<50 pc), outer regions (>50 pc), and general trends over time. Increases (\(\uparrow\)), decreases (\(\downarrow\)), and no change (\(\rightarrow\)) are indicated for the CI/CO, CII/CO, and CI/CII ratios. The table highlights that the CI/CO and CI/CII ratios generally decrease over time, while the CII/CO ratios show an increasing trend. Notably, the AGN model exhibits lower median values for CI/CO, CII/CO, and CI/CII before 8 Myr, with a reversal in the CI/CO ratio at 8 Myr, suggesting changing abundances influenced by AGN activity.
    }
    \begin{tabular}{|l|c|c|c|c|c|c|l|}
        \hline
        \textbf{Species/Ratio} & \multicolumn{3}{c|}{\textbf{AGN}} & \multicolumn{3}{c|}{\textbf{NoAGN}} & \textbf{Description} \\ \hline
        & \textbf{Inner} & \textbf{Outer} & \textbf{General} & \textbf{Inner} & \textbf{Outer} & \textbf{General} \\ \hline
        \textbf{CI}            & \(\downarrow^a\)          & \(\uparrow^b\)        & \(\downarrow\)         & \(\uparrow\)        & \(\downarrow\)     & \(\downarrow\)     & $a$: \(\uparrow\) at t=8 Myr; $b$: \(\rightarrow\) at t < 6 Myr \\ \hline
        \textbf{CII}           & \(\downarrow\)        & \(\uparrow^a\)      & \(\downarrow^b\)       & \(\uparrow\)        & \(\downarrow^a\)     & \(\downarrow\)     & $a$: \(\rightarrow\) at t < 8 Myr; $b$: \(\uparrow\) at inner t = 7,8 Myr \\ \hline
        \textbf{CO}            & \(\downarrow^a\)          &  \(\uparrow\)        & \(\downarrow^b\)         &  \(\uparrow\)        & \(\downarrow\)        & \(\downarrow^b\)     & $a$: \(\uparrow\) at t=7,8 Myr; $b$: \(\uparrow\) at inner t = 7,8 Myr\\ \hline
        \textbf{CI/CO}   & \(\downarrow^a\)        & \(\downarrow^b\)        & \(\downarrow^a\)         & \(\uparrow\)     & \(\uparrow\)         & \(\downarrow\)      & $a$: \(\uparrow\) at t=8Myr; $b$: \(\uparrow\) at t=7,8 Myr \\ \hline
        \textbf{CII/CO}  & \(\downarrow\)        & \(\downarrow\)      & \(\uparrow\)       & \(\uparrow\)        & \(\uparrow\)       & \(\uparrow^a\)     & $a$: \(\downarrow\) inner at t=8Myr \\ \hline
        \textbf{CI/CII}  & \(\downarrow^a\)        & \(\uparrow\)        & \(\downarrow^b\)         & \(\uparrow\)        & \(\downarrow\)       & \(\downarrow\)      &$a$: \(\uparrow\) at t=7,8 Myr; $b$: \(\uparrow\) inner and at t = 7,8 Myr \\ \hline
    \end{tabular}
    \label{tab:species_ratios}
\end{table*}

 \begin{figure}
	\centering
 \includegraphics[width=0.99\linewidth]{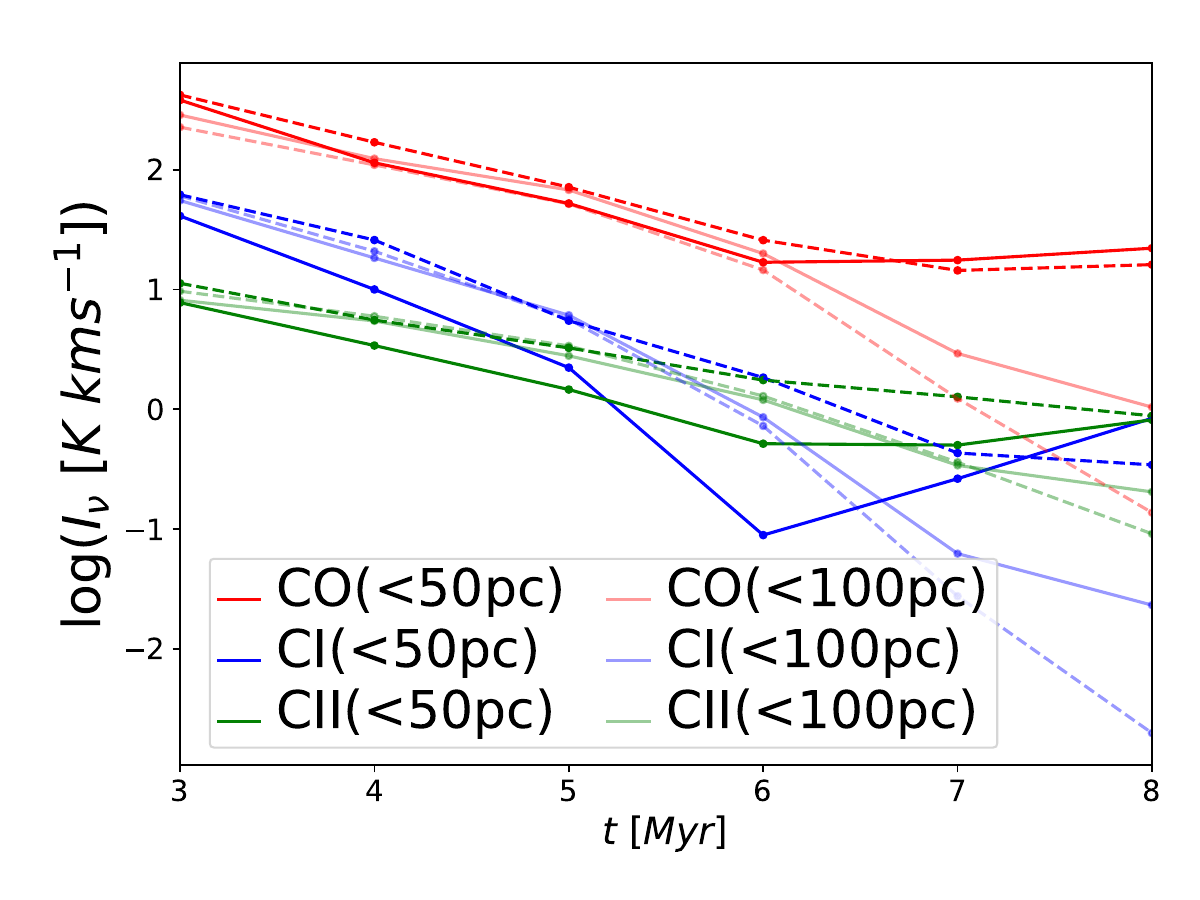}
 \includegraphics[width=0.99\linewidth]{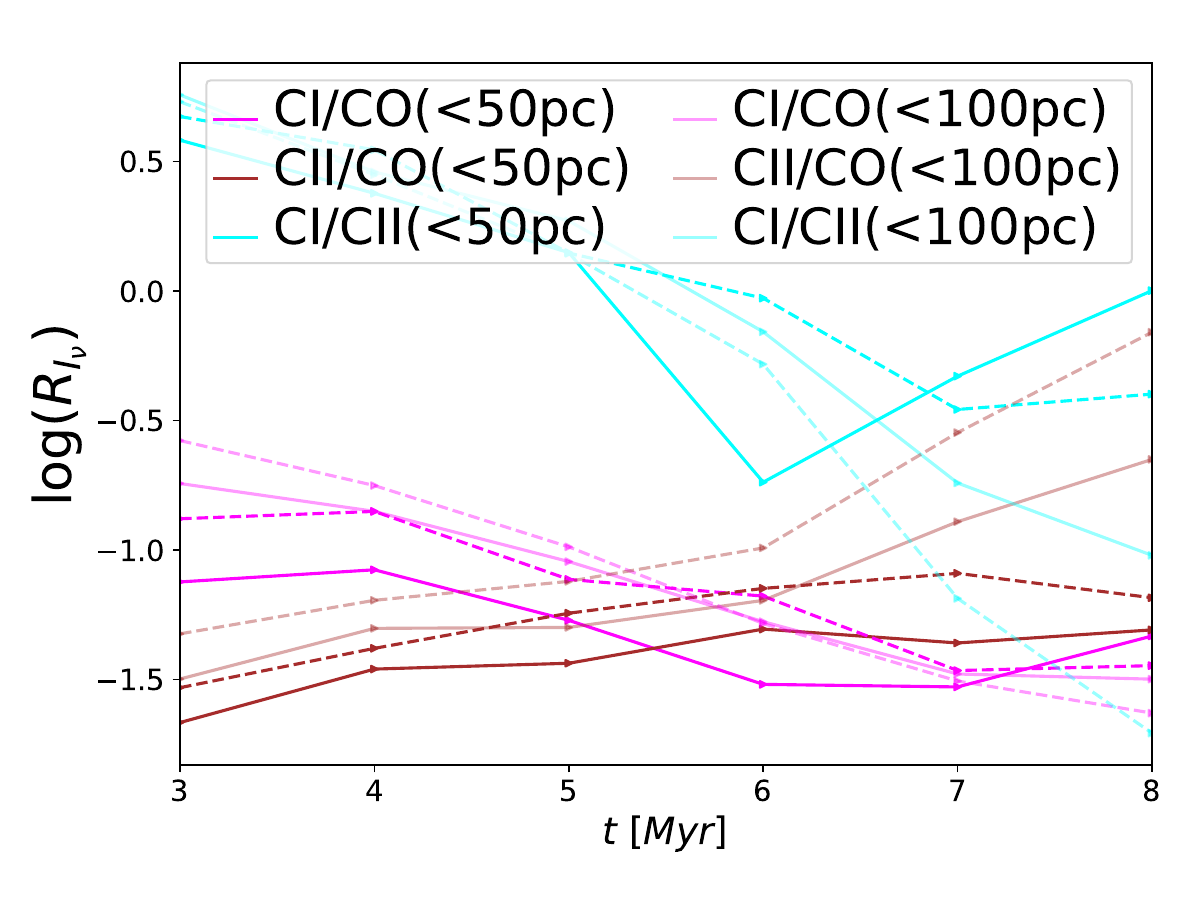}
	\caption{
     Top panel: Median integrated intensity measurements for radii less than 50 pc and 100 pc from the center of the disc, including species such as CO, CI, and CII, collected over a time span of 3 to 8 Myr. The data reveal a declining trend in intensity over time, more pronounced at larger radii (<100 pc), indicating that intensity decreases as distance from the center increases in both AGN and NoAGN models. In the inner and outer regions, median intensities of CI, CII, and CO are lower in the AGN model for time scales below 7 Myr. However, after 7 Myr, the inner region of the AGN model shows an increase in median CO intensity, followed by increases in CI and CII at 8 Myr, suggesting AGN influence on the distribution and abundance of these species. Bottom panel: Median integrated intensity ratios for CI/CO, CII/CO, and CI/CII for radii smaller than 50 pc and 100 pc. The trends indicate a gradual decrease in CI/CO and CI/CII ratios over time, reflecting a diminishing abundance of CI relative to CII. Notably, the CI/CII ratio increases after 6 Myr for radii below 50 pc. CII/CO ratios exhibit a generally increasing pattern over time. In the inner regions, AGN models show lower median values for CI/CO, CII/CO, and CI/CII for time scales below 8 Myr compared to NoAGN models, with a notable increase in CI/CO at 8 Myr for the AGN model. Overall, the NoAGN model shows a higher abundance of CI relative to CO before 7 Myr, a trend that reverses at 8 Myr in the inner regions. 
 }
	\label{fig:Median_50_100}
\end{figure}

 \begin{figure}
	\centering
 \includegraphics[width=0.8\linewidth]{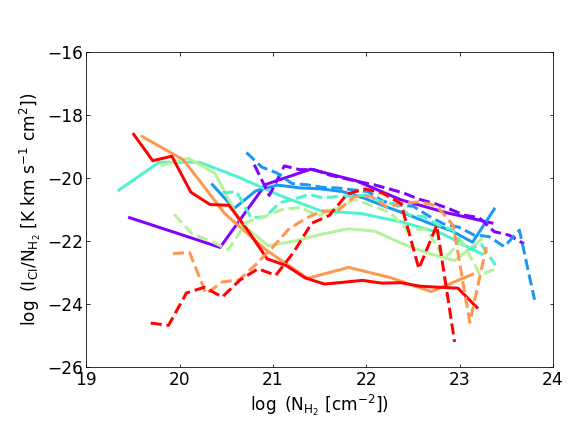}
 \includegraphics[width=0.8\linewidth]{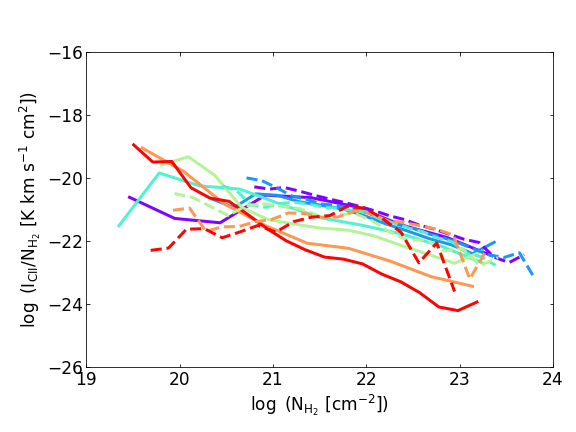}
 \includegraphics[width=0.8\linewidth]{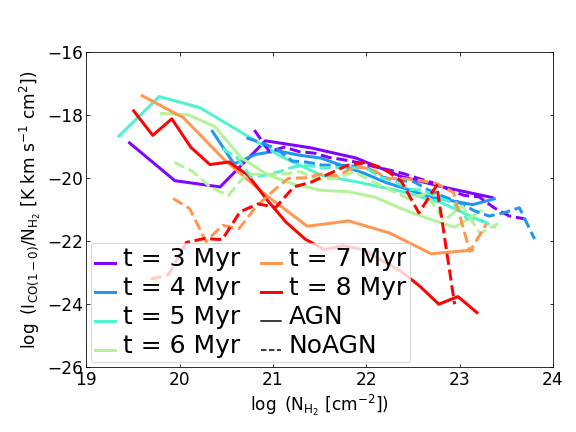}
  \includegraphics[width=0.8\linewidth]{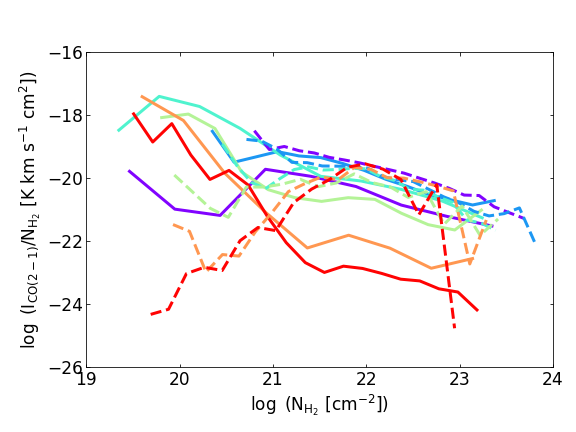}
	\caption{
     Intensity ratios of CO(2-1), CO(1-0), CI, and CII to the $\rm H_2$ column density ($N_{\rm H_2}$) as a function of $N_{\rm H_2}$ for time scales between 3 and 8 Myr (colors) for AGN (solid lines) and NoAGN (dashed lines) models. The AGN model shows a "CO-dark" region at later stages (t = 7, 8 Myr), where CO intensity ($I_{CO}$) diminishes relative to $N_{\rm H_2}$ due to AGN-induced disruption of molecular gas. This affects lower-excitation CO transitions more than higher-J transitions. Conversely, the $\rm I_{CI}/N_{H_2}$ and $\rm I_{CII}/N_{H_2}$ ratios may increase, indicating enhanced atomic and ionized carbon abundance as CO dissociates, reflecting altered conditions in the CND due to AGN feedback. 
    }
	\label{fig:Intensity_NH2}
\end{figure}

 \begin{figure}
	\centering
 \includegraphics[width=0.45\linewidth]{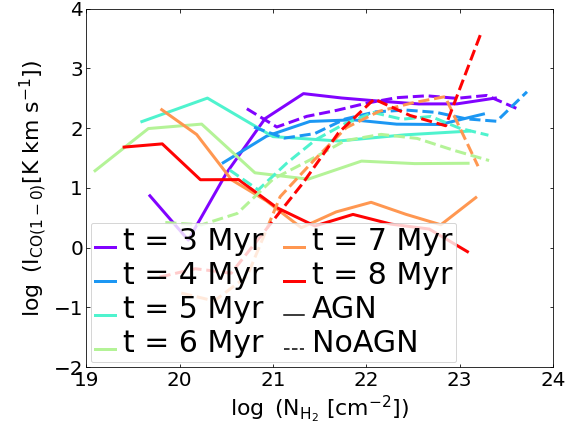}
 \includegraphics[width=0.45\linewidth]{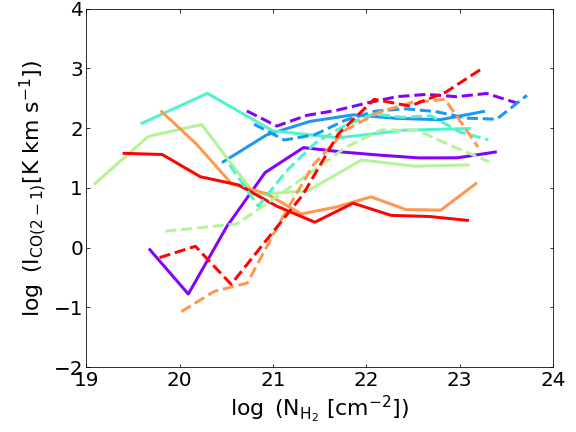}
 \includegraphics[width=0.45\linewidth]{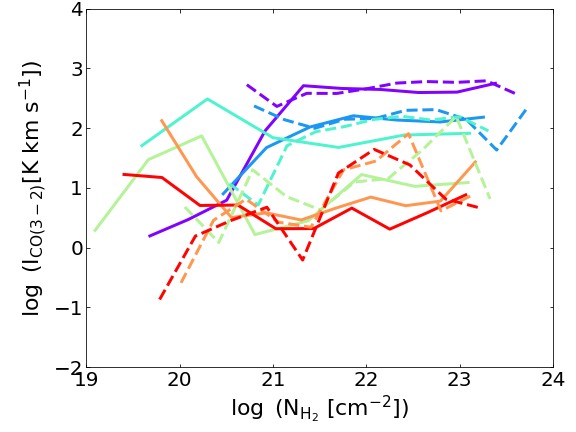}
\includegraphics[width=0.45\linewidth]{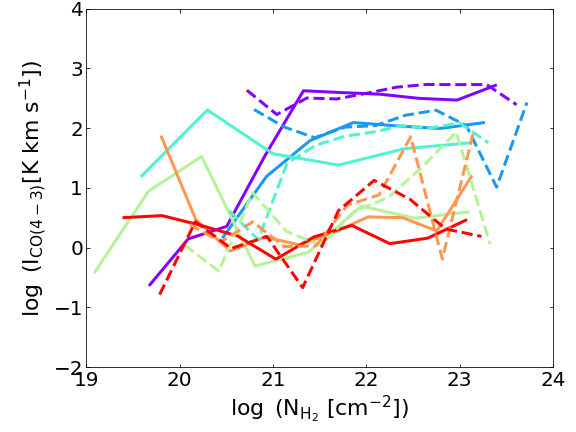}
 \caption{  Relationships between CO J-line intensities and the H$_2$ column density ($N_{\rm H_2}$) for time scales between 3 and 8 Myr (colors) for AGN (solid lines) and NoAGN (dashed lines) models. In typical molecular clouds, lower-J CO transitions (e.g., CO(1-0) and CO(2-1)) exhibit a linear relationship with $N_{\rm H_2}$. However, in the circumnuclear disc (CND) influenced by AGN feedback, these relationships deviate, particularly in the CO-dark region, where CO(1-0) and CO(2-1) intensities show significant depletion relative to $N_{\rm H_2}$. This indicates disruption and dissociation of molecular gas due to AGN mechanical feedback, which primarily affects lower-excitation CO transitions. In contrast, higher-J transitions (e.g., CO(3-2) and CO(4-3)) show less depletion or even an increase, reflecting selective heating and excitation of the remaining molecular gas.
}
	\label{fig:Intensity_CO_J_NH2}
\end{figure}

\subsection{The Median Integrated Intensity}

The top panel of Figure \ref{fig:Median_50_100} displays the median integrated intensity measurements from radii less than 50 pc and 100 pc from the center of the disc. These measurements, which include species such as CO, CI, and CII, were collected over a time span of 3 to 8 Myr. Overall, the data indicate a declining trend in intensity over time, with this decline being more pronounced at larger radii (<100 pc). This suggests that the intensity of these species decreases as we move further from the center of the disc in both the AGN and NoAGN models.
In both the inner and outer regions, the median intensities of CI, CII, and CO are lower in the AGN model for time scales below 7 Myr compared to the NoAGN model. However, in the AGN model, after 7 Myr, the inner region shows an increase in the median intensity of CO, followed by similar increases for the inner and outer regions of CI and CII at 8 Myr. This suggests that the presence of an AGN influence the distribution and abundance of these atomic species within the disc.

The bottom panel of Figure \ref{fig:Median_50_100} illustrates the median integrated intensity ratios obtained from radii smaller than 50 pc and 100 pc from the center of the disc for CI/CO, CII/CO, and CI/CII in both the AGN and NoAGN models. Overall, the trends reveal a gradual decrease over time for the CI/CO and CI/CII ratios, indicating a diminishing abundance of CI relative to CII as the system evolves. Notably, the CI/CII ratio shows an increase after 6 Myr for radii smaller than 50 pc.
Additionally, the CII/CO ratios exhibit relatively increasing patterns over time. In the inner regions, the AGN models display lower median values for CI/CO, CII/CO, and CI/CII for time scales below 8 Myr compared to the NoAGN model. However, there is an increase in the CI/CO ratio at 8 Myr for the AGN model. As discussed in Section \ref{sec:profile}, the CI/CO ratio is higher in the NoAGN model compared to the AGN model before 7 Myr, suggesting a more prominent presence of CI relative to CO in the absence of AGN activity. This trend reverses at t = 8 Myr in the inner regions (< 50 pc). Furthermore, the CII/CO ratios are predominantly higher for the NoAGN model, indicating a relatively greater abundance of CII compared to CO when an AGN is not present. Interestingly, the AGN model shows an increasing trend for CI after 6 Myr.

 Table \ref{tab:species_ratios} summarizes the trends of increases and decreases in the inner and outer regions for various species and ratios in both the AGN and NoAGN models. It categorizes trends in median intensities of CI, CII, CO, and their ratios by inner regions ($<$50 pc), outer regions ($>$50 pc), and overall trends over time. Changes are indicated as increases ($\uparrow$), decreases ($\downarrow$), or no change ($\rightarrow$) for the CI/CO, CII/CO, and CI/CII ratios. The table highlights that the CI/CO and CI/CII ratios generally decline over time, while the CII/CO ratios tend to increase. The AGN model shows lower median values for CI/CO, CII/CO, and CI/CII before 8 Myr, with a reversal in the CI/CO ratio at that time, suggesting that AGN activity influences changes in abundances.

\subsection{CO-dark regions}

In Figure \ref{fig:Intensity_NH2}, the intensity ratios of CO(2-1), CO(1-0), CI, and CII to the $\rm H_2$ column density ($N_{\rm H_2}$) reveal the relative abundance and behavior of these carbon-bearing species within the interstellar medium and molecular clouds.  The above Figure illustrates that in the NoAGN model, during the early timescale (t < 6 Myr), the column density remains above \( 10^{21} \, \text{cm}^{-3} \), with CO, CI, and CII decreasing as $N_{\rm H_2}$ increases. Starting at t = 6 Myr, the intensities of CO, CI, and CII rise with $N_{\rm H_2}$ for column densities below \( 10^{21} \, \text{cm}^{-3} \), with CO and CI showing a steeper increase at later times (t = 7, 8 Myr) before declining for higher column densities. This behavior suggests that the NoAGN environment initially limits carbon species formation but becomes more favorable as conditions evolve, particularly at lower densities. In contrast, the AGN model exhibits different trends. Here, CO consistently decreases across all timescales, with a more pronounced decline after t > 7 Myr, indicating that the AGN environment may suppress CO formation more effectively over time. CI also shows a gradual decrease, with a sharp decline at $N_{\rm H_2}$ below \( 10^{21} \, \text{cm}^{-3} \) before stabilizing. Similarly, CII decreases across all timescales with increasing $N_{\rm H_2}$, reflecting the AGN's impact on ionization processes.

In the context of the CND around an AGN, the shape and trends of these ratios can be affected by the AGN's mechanical feedback. A notable feature in the later stages (t = 7,8 Myr) of the AGN model is the emergence of a "CO-dark" region, where the intensity of CO emission ($I_{CO}$) is diminished relative to the H$_2$ column density compared to the NoAGN model. This CO-dark region likely arises from the AGN's effects, which can disrupt and dissociate the molecular gas, resulting in a decreased CO abundance relative to other carbon-bearing species, particularly evident at $t = 8$ Myr. This disruption mainly impacts lower-excitation CO transitions, while higher-J CO transitions (e.g., CO(2-1)) may show a less pronounced depletion or even an increase in their intensity ratios, as the mechanical feedback can also heat and excite the remaining molecular gas.

 In contrast, as shown in the top two panels of Figure \ref{fig:Intensity_NH2}, the $\rm I_{CI}/N_{H_2}$ and $\rm I_{CII}/N_{H_2}$ ratios exhibit distinct behaviors.  The scatter in the $\rm I_{CI}/N_{H_2}$ ratios is approximately 2 dex larger than that observed for the $\rm I_{CII}/N_{H_2}$ ratios across different timescales, indicating that the CI emission responds more variably to environmental changes. Furthermore, AGN activity becomes increasingly efficient at reducing the CI emission at later timescales in the $\rm I_{CI}/N_{H_2}$ ratio compared to the NoAGN model. In contrast, the $\rm I_{CII}/N_{H_2}$ ratio in the AGN scenario exhibits lower scatter, influenced by local conditions, when compared to the NoAGN model.
This could indicate the enhanced abundance of atomic and ionized carbon (CO gets dissociated) compared to molecular hydrogen, as the mechanical feedback from the AGN alters the chemical and physical conditions in the CND \citep[the increase in atomic carbon in CO dark regions has been shown by various studies;][]{Bisbas2015,Franeck2018}. 

Figure \ref{fig:Intensity_CO_J_NH2} also shows that the relationships between these CO J-line intensities and the H$_2$ column density ($N_{\rm H_2}$) can reveal important information about the CO excitation and the impact of AGN feedback.
In a typical molecular cloud environment, one would expect to see a certain pattern in the CO J-line intensities as a function of $N_{\rm H_2}$. The lower-J CO transitions, such as CO(1-0) and CO(2-1), would typically exhibit a linear or near-linear relationship with $N_{\rm H_2}$, reflecting the relatively uniform CO excitation conditions.
However, in the CND influenced by the mechanical feedback from the AGN, the CO J-line intensity versus $N_{\rm H_2}$ relationships deviate from this expected pattern, particularly within the CO-dark region. The CO(1-0) and CO(2-1) line intensities show a depletion relative to the H$_2$ column density, indicating a disruption and dissociation of the molecular gas by the AGN's mechanical feedback, which preferentially affects the lower-excitation CO transitions.  As illustrated in Figure \ref{fig:Intensity_CO_J_NH2}, the CO lines in AGN-affected regions compared to NoAGN regions demonstrate a depletion. This depletion suggests a reduction in the abundance of CO molecules in areas influenced by AGN activity. We will conduct further analysis on the CO intensity ratios in our future study.
In contrast, the higher-J CO transitions, such as CO(3-2) and CO(4-3), exhibit a less pronounced depletion or even an increase in their intensity versus $N_{\rm H_2}$ relationships. This can be attributed to the AGN feedback selectively heating and exciting the remaining molecular gas, leading to a relative enhancement of the higher-J CO lines compared to the lower-J transitions.

\section{Discussion}
\label{subsec:sum}

In this study, we utilized the HDGAS hydrodynamic simulation \citep{Raouf2023} to investigate the impact of active galactic nucleus (AGN) feedback on the interstellar medium by analyzing various atomic to molecular line ratios, including CI/CO, CII/CO, and CI/CII, as well as the abundances of different species, to explore the transition from neutral atomic gas to molecular gas in models with and without AGN influence. Comparing the AGN and NoAGN models, we show significant differences in the intensity maps and radial profiles of CI, CII and CO and their ratios. The results show that the CI/CO ratio serves as a useful indicator of the transition from atomic to molecular gas. 

The probability distribution functions (PDFs) of these species demonstrate the influence of an AGN on the relative abundances within various scales of the circumnuclear disc (50 pc and 100 pc). The CI PDF indicates a time-dependent increase in CI density specifically for the AGN model. Meanwhile, the CO PDFs reveal that the AGN model exhibits higher density tails at both the 50 pc and 100 pc scales of the disk, highlighting the AGN's impact on the distribution of CO as well \citep[as shown in][regarging positive feedback from AGN]{Raouf2023}. The AGN model exhibits  elevated levels of CI and CII gas abundances across various time scales.  We observe notable differences in carbon species abundance between the AGN and NoAGN models, particularly in the region \( r < 50 \, \text{pc} \). Both models exhibit a peak logarithm of CO abundance relative to hydrogen density \( \left( \frac{n_{\text{CO}}}{n_{\text{H}}} \right) \) at \(-3.7\), indicating similar conditions for CO formation. However, the AGN model shows a higher peak for CI at \(-5\), compared to \(-5.5\) for NoAGN, suggesting that the AGN enhances CI abundance through increased heating and photodissociation. CII peaks at \(-8\) in the AGN model and \(-8.3\) in the NoAGN, reflecting greater ionization in the AGN environment. At \( r < 100 \, \text{pc} \), the CO probability distribution function (PDF) peaks at \(-4\) for NoAGN and \(-3.5\) for AGN, indicating higher CO abundance in the AGN model, likely due to shock heating and enhanced star formation. These results underscore the significant influence of an active galactic nucleus on the chemical pathways and abundances of carbon species.

In the NoAGN model, the intensity maps show higher values of CI and CO at radii below 50 pc and time scales under 7 Myr, suggesting a more pronounced interface between ionized and molecular gas, likely due to local star formation processes. In contrast, the AGN model reveals increased CI and CO intensities in the outer regions (greater than 50 pc) at time scales exceeding 7 Myr, indicating that AGN feedback contributes to additional heating and excitation of the remaining molecular gas.
The AGN model also exhibits a higher CI/CO ratio at larger radii at later time points, reflecting the AGN's influence on photodissociation and the chemical composition of the gas. Meanwhile, the NoAGN model shows a dominant CI/CO ratio at smaller radii, highlighting a different balance of physical processes.
This aligns with observations of galaxies like NGC 1808 and NGC 7469, which show higher CI/CO intensity ratios in their outer regions. While our simulations predict CO emissions extending to larger radii, the observed upward tails in CI versus CO ratios are mainly due to detection limits. Overall, our models indicate that CO is higher in the AGN model, particularly in outer regions ($>$ 50 pc) and inner region at later times ($>$ 7 Myr).  Further, the observed trends are limited to the range \( 1 < \log(I_{\text{CO}}) < 3 \), while simulations extend to lower CO values, \( -1 < \log(I_{\text{CO}}) < 3 \), suggesting that conditions for lower CO abundance exist beyond the observed limits. The scatter in CI intensity brightness is approximately \(-2 \, \text{dex}\), indicating stability, whereas the CI/CO ratio shows greater variability at about 
$-2.5$ dex. Lower CI/CO ratios at distances 
\( r > 200 \, \text{kpc} \) imply less favorable conditions for CI formation due to lower densities and reduced star formation activity. Additionally, time evolution contributes to these trends, highlighting the dynamic interplay between carbon species and the influence of galactic structure on chemical evolution.

The higher CII emission in the NoAGN model's core suggests that the absence of AGN feedback helps preserve dense CII-emitting gas. The overall decrease in CII emission in outer regions may result from star formation and other galaxy-scale processes. The CI/CII ratio profile reflects a complex interaction between the AGN and NoAGN cases, with the AGN model initially showing a lower CI/CII ratio in the central region but a higher ratio in outer regions over time. This suggests that the AGN enhances CI production relative to CII in these areas due to its influence on ionization and thermal conditions.

  In the NoAGN model, column density remains above \( 10^{21} \, \text{cm}^{-3} \) during the early timescale (t < 6 Myr), with CO, CI, and CII decreasing as $N_{\rm H_2}$ increases. However, after t = 6 Myr, their intensities rise with $N_{\rm H_2}$ at lower column densities before declining again. This indicates a shift toward more favorable conditions over time. Conversely, in the AGN model, CO consistently decreases across all timescales, especially after t > 7 Myr, reflecting stronger suppression. CI and CII also decline with increasing $N_{\rm H_2}$, emphasizing the AGN's significant impact on the chemical dynamics of carbon species. Additionally, our findings highlight the presence of a "CO-dark" region in the AGN model, where lower-J CO transitions (e.g., CO(1-0) and CO(2-1)) are significantly depleted relative to H$_2$ column density. This depletion is likely a result of the AGN's mechanical feedback at later time scale, which disrupts and dissociates CO molecular gas. In contrast, higher-J CO transitions may show less pronounced depletion or even an increase in intensity, as the AGN can lead to heating and excitation of the remaining gas.

It is important to note that the specific impact of the AGN on the integrated intensity maps and radial profiles of all molecules and atomic species in this study can vary depending on the properties of the AGN, the molecular gas environment, and the specific stage of AGN activity. Observational studies and modeling efforts are essential for characterizing the AGN's influence and understanding the underlying physical processes.

\section*{Acknowledgments} \label{sec:acknow}
MR and SV acknowledge support from the European Research Council (ERC) Advanced grant MOPPEX 833460. MR would like to acknowledge SurfSARA Computing and Networking Services for their support (EINF-5315). MR would like to express his heartfelt appreciation to EuroSpaceHub and LUNEX EuroMoonMars Earth Space Innovation for their generous funding and unwavering support.   

Data manipulation and analysis were performed using the Pandas Python package \citep{reback2020pandas}. Pandas is a powerful open-source library that provides efficient data structures and data analysis tools, making it a popular choice for handling structured data in the Python programming language.

\section*{Data availability}
Data from the simulations is based on the output from publicly available GIZMO \citep{Hopkins2015} code at the following repository: \href{https://bitbucket.org/phopkins/gizmo-public/src/master/}{https://bitbucket.org/phopkins/gizmo-public/src/master/}  generated on the ICs described in section \ref{sec:simulation}. \textit{The data underlying this article will be shared on reasonable request to the corresponding author}.
In the Data Availability section, you can also point to the public website for the CHIMES code, e.g. "The simulations in this paper also used the CHIMES non-equilibrium chemistry and cooling code which is publicly available from this webpage: https://richings.bitbucket.io/"

\bibliographystyle{mnras}
\bibliography{bibliography.bib}
%\bsp

\label{lastpage}
\end{document}